\documentclass[fa4paper,11pt]{article}
\pdfoutput=1 

\usepackage{jheppub} 

\usepackage[T1]{fontenc} 
\usepackage{subfigure}
\usepackage{booktabs}
\usepackage{physics}
\usepackage{feynmf}
\usepackage{tikz}

\title{\boldmath Entanglement, Renormalization and Effective Field Theories}

\author[a,b]{Bingzheng Han}
\author[a]{and Ratindranath Akhoury}
\affiliation[a]{Leinweber Center for Theoretical Physics,\\Randall Laboratory of Physics, Department of Physics,\\University of Michigan, Ann Arbor, MI 48109, U.S.A}
\affiliation[b]{Joseph Henry Laboratories, Princeton University\\ Princeton, NJ 08544, U.S.A}

\emailAdd{hbzzz@princeton.edu}
\emailAdd{akhoury@umich.edu}

\abstract{We develop the idea that renormalization, decoupling of heavy particle effects from low energy physics and the construction of effective field theories 
are intimately linked to the momentum space entanglement of disparate modes of an interacting quantum field theory. Using unitary transformations to decouple these modes at the perturbative level, we show in a scalar field theoretical model with light and heavy fields, how renormalization may be consistently implemented and how the low energy effective field theory can be constructed. We also obtain a renormalization group equation in this framework and apply it to the scalar field theoretical model.}

\begin{document} 
\maketitle
\flushbottom

\section{Introduction}
\label{sec:intro}

Entanglement is ubiquitous in any quantum theory. In a free field theory 
the different momentum modes are not entangled. However, the introduction of interactions causes the entanglement, in particular, of the low momentum modes with the 
inaccessible high energy ones. In experiments only the low energy or larger wavelength modes are accessible and renormalization can be thought of as a procedure to disentangle the high energy modes from those of low energy at the same time incorporating the effects of the former in a modified effective theory at low energies. In the usual Wilsonian approach \citep{Wilson:1971bg}\cite{Wilson:1971dh}\cite{Polchinski:1983gv}, the high energy modes are integrated out and in this way we arrive at a low energy effective action. An alternative viewpoint, that we discuss here, is to directly address the entanglement and by a series of unitary transformations decouple the
low and high energy modes. The effective low energy Hamiltonian is then obtained by projecting onto the ``high energy vacuum", i.e., the low energy subspace where there are no modes of heavy masses or of momenta larger than some cut-off scale which can appear as external states. In this paper, we discuss renormalization, decoupling of heavy mass states \cite{Appelquist:1974tg} and the construction of effective field theories, \cite{Georgi:1994qn}\cite{Manohar:1996cq}\cite{Rothstein:2003mp}, all from this perspective. Not only do we set up the general formalism but  also exemplify the procedure by explicit examples up to the one-loop level and to order $1/M^2$, where $M$ is the heavy mass scale. Our results are in agreement with those obtained by the standard methods (see for example, \cite{Rothstein:2003mp}) showing that such a program can be successfully implemented,and thereby providing another way to construct low energy effective theories. In terms of an extended program, this paper is a first step in exploring how entanglement measures may be used in general to address problems in quantum field theory like the correlations between different momentum scales and in this way provide another window into renormalization and related phenomenon.\\

Envisioning the renormalization process as removing the entanglement between the low and high momentum modes through unitary transformations has a clear physical significance in the Hamiltonian or Schrodinger framework which we adopt in this paper. This framework is not manifestly covariant in the intermediate stages and has seldom been used for practical calculations, though it has been studied at a formal level, for example, in \cite{Symanzik:1981wd}, \cite{Pi:1987df}, \cite{Minic:1994ff}.In particular, in spite of the great successes of the effective field theory approach, rarely has work been done in a Hamiltonian framework. Ours is a straightforward and direct attempt where we formally connect the Hamiltonian approach to the standard one by mapping the calculations in this framework to the usual Feynman diagram calculations of S matrix elements and extend dimensional regularization techniques to perform loop calculations. Methods to evaluate certain unusual integrals encountered in this approach are discussed in appendices. \\

The study of renormalization through similarity transformations has a long history, though most of the previous works do not reference the crucial connection to momentum space entanglement. In \cite{Glazek:1994qc} similarity transformations were introduced to control divergences in light-front field theory. An exact renormalization group equation similar in appearance to the one in section (\ref{sec:Renormgroup}) of this paper was obtained there. In \cite{Gubankova:1997mq}, a similar set of Hamiltonian flow equations were obtained but it was tailored to address problems in many body theory where no divergent renormalization is necessary. The motivation of these two references was also very different from the one pursued here. The analysis presented in sections, (2) and (3) are closest in spirit to the perturbative Hamiltonian renormalization of a scalar theory with purely quartic interactions discussed in \cite{Alexanian:1998wu}. In this paper our emphasis is on renormalization from the perspective of momentum space entanglement and we also similarly study the decoupling of heavy particle effects and the construction of effective field theories in the context of a model involving both heavy and light fields. In addition, we also address the problem of renormalization group flow in this approach. In a more recent paper, \cite{Balasubramanian:2011wt} momentum space entanglement and renormalization in a quantum field theory  has been specifically addressed. There, a relationship is obtained between the Wilsonian effective action and the density matrix (with an entanglement entropy) describing the infrared degrees of freedom of the theory. Though there are points of contact, our approach is more direct and there is little connection with the techniques or the results of \cite{Balasubramanian:2011wt} .\\

The paper is organized as follows.
In section (\ref{sec:Theroy}) we discuss our general approach to a perturbative realization of the disentangling high and low momentum modes by means of unitary transformations on the states. We discuss the projection of the Hamiltonian of a theory onto the low energy subspace and the procedure followed in the rest of the paper for renormalization and construction of effective field theories .
In section (\ref{sec:Calc}) we consider a scalar field theoretical model with heavy and light fields and explicitly construct the unitary transformation that shows clearly how renormalization and decoupling works. 
In particular, we construct the light particle two and four point functions up to order $\frac{1}{M^2}$ in the heavy mass expansion and at one loop order.
In section (\ref{sec:EFT}) we extend the previous construction to obtain an effective field theory of the light fields alone and make connection with previous work based on conventional methods. In section (\ref{sec:Renormgroup}) we check the consistency of our approach by setting up a renormalization group equation and discuss an evaluation of the $\beta$ function in a scalar theory with only quartic interactions.
We conclude with a discussion of these results in section (\ref{sec:discussion}). Certain technical details of the calculations, particularly those encountered in section (\ref{sec:Calc}) are relegated to appendices.

\section{Perturbative Decoupling, Renormalization and Matching of Hamiltonian Operators}
\label{sec:Theroy}
\subsection{Decoupling with Unitary Transformations}
\label{subsec:ch21}
The subject of decoupling in Effective Field Theory has been studied extensively for the past many decades. The decoupling theorem states that if the low energy effective theory is renormalizable, and a physical renormalization scheme has been applied, then all effects due to heavy particles will appear as changes to couplings or are suppressed as $\frac{1}{M^n}$, where M is the mass of the heavy particle. As we discussed in the introduction, an alternative way to consider the decoupling is to introduce a series of unitary transformations to decouple high energy and low momentum modes and then look at the low energy part of the spectrum.
The Hamiltonian framework is best suited to study decoupling and renormalization from this perspective. In this fixed time approach, consider the action of a unitary transformation $\omega$ on the states of a theory,
\begin{equation}
    \ket{\Psi(\mu)} = \omega^{\dagger}(\mu)\ket{\Psi},
\end{equation}
under which the Hamiltonian transforms as $H'=\omega^{\dagger}H\omega$.
We would like to use this unitary transformation to disentangle the momentum states above a scale $\mu$ from the low energy ones. However, this exact diagonalization procedure, in general, is impossible at present and we have to be content with a less ambitious, perturbative approach, where we identify a low energy subspace by projecting the unitarily transformed Hamiltonian on to a state which acts as a vacuum for high energy particles.
Let us label this state as $\ket{0_{H}}$ which satisfies,
\begin{equation}
    \begin{aligned} a_{H}\ket{0_H}=0,
    \end{aligned}
\end{equation}
where, $a_{H}$ is the annihilation operator of high energy modes. Thus in this approach the job of the unitary transformation, order by order is to remove the terms in the full Hamiltonian which will perturb the high energy vacuum structure. These are terms containing only high energy creation operators. Since Hamiltonian operator is Hermitian, $\omega$ will inevitably cancel terms containing only high energy annihilation operators as well. This condition allows us to identify the unitary transformations perturbatively and the transformed Hamiltonian with this $\omega$ when projected on to the high energy vacuum will give us what we refer to as the decoupled Hamiltonian at low energy. The decoupled Hamiltonian encodes the effects of the high momentum modes on the low energy physics. The precise way this is accomplished will now be discussed. A similar method was used in a related context in \cite{Alexanian:1998wu}.\\

\indent Let's consider the Hamiltonian $H$ of a full theory, and denote  $H_{eff}(\mu)$ as the effective Hamiltonian defined at an energy scale $\mu$. $H_{eff}(\mu)$ will generate the same physical results i.e. S-matrix elements for all the physical processes that do not involve momenta greater than $\mu$. We can view $H_{eff}(\mu)$ as the projection of the full theory onto the low energy subspace:
\begin{equation}
\label{eq:x}
    \begin{aligned} 
    H_{eff}(\mu)=P(\mu)HP(\mu),
    \end{aligned}
\end{equation}
where P($\mu$) is the projection operator at energy scale $\mu$.\\
As discussed, at least perturbatively we can decouple low energy modes from high energy ones using a series of unitary transformations and thus construct the high energy vacuum and obtain the low energy subspace. Let $H_{decoupled}$ denote the decoupled Hamiltonian at low energy:
\begin{equation}
    \label{eq:x}
    \begin{aligned} H_{decoupled}=\bra{0_{H}} \omega^{\dagger} H \omega \ket{0_{H}},
    \end{aligned}
\end{equation}
In general, the $\omega$ here is a product series of unitary transformations.  Furthermore, we will normal order with respect to high energy vacuum. It's worth noting here that the $H_{decoupled}$ so calculated is an intermediate step towards the calculation of the physical effective Hamiltonian $H_{eff}$. However, as we will see later its components have an important physical meaning regarding renormalization and further it will also be involved in the matching process to get the physical effective Hamiltonian $H_{eff}$.\\
\indent Let's break $\omega$ into a product series:
\begin{equation}
    \label{eq:x}
    \begin{aligned} \omega=\omega_{0}\omega_{1}\omega_{2}...\omega_{n}...
    \end{aligned}
\end{equation}
Each $\omega_{i}$ partially diagonalizes the Hamiltonian to a given order $\sim \frac{1}{\mu}$, $\mu$ is the cut-off energy scale. We can decompose the full Hamiltonian as:
\begin{equation}
    \label{eq:x}
    \begin{aligned} H=H_{1}+H_{2}+H_{A}+H_{B},
    \end{aligned}
\end{equation}
where $H_{1}$ only contains low energy modes, $H_{2}$ is the free part for high energy modes, $H_{A}$ contains terms that only have high energy annihilation or creation operators and $H_{B}$ is whatever left. For simplicity, we can set $H_{1}$ to be $O(1)$ in energy , and the other three terms of $O(\mu)$.\\
\indent Let's consider the following:
\begin{equation}
    \label{eq:x}
    \begin{aligned} &\omega_{0}^{\dagger}\left(H_{1}+H_{2}+H_{A}+H_{B}\right)\omega_{0}\\
    =&e^{-i\Omega_{0}}\left(H_{1}+H_{2}+H_{A}+H_{B}\right)e^{i\Omega_{0}}\\
    =&H_{1}+H_{2}+H_{A}+H_{B}+i[H_{1},\Omega_{0}]+i[H_{2},\Omega_{0}]+i[H_{A},\Omega_{0}]+i[H_{B},\Omega_{0}]...
    \end{aligned}
\end{equation}
We want to eliminate $H_{A}$ by choosing $\Omega_{0}$ such that 
\begin{equation}
    \label{eq:x}
    \begin{aligned} i[H_{2},\Omega_{0}]+H_{A}=0.
    \end{aligned}
\end{equation}
This is our decoupling condition at $O(\mu)$, and since both $H_{2}$ and $H_{A}$ are of $O(\mu)$, we can deduce that $\Omega_{0} \sim O(1)$. Although we cancel out $H_{A}$, we create a new term $i[H_{1},\Omega_{0}]$ of order $\sim O(1)$ that contains only annihilation or creation operators and in order to eliminate this new term, we need to introduce the next unitary operator $\omega_{1}=e^{i\Omega_{1}}$ at $O(\frac{1}{\mu})$. Then the Hamiltonian becomes
\begin{equation}
    \label{eq:x}
    \begin{aligned} &e^{-i\Omega_{1}}e^{-i\Omega_{0}}\left(H_{1}+H_{2}+H_{A}+H_{B}\right)e^{i\Omega_{0}}e^{i\Omega_{1}}\\
    =&H_{1}+H_{2}+H_{B}+i[H_{1},\Omega_{0}]+i[H_{A},\Omega_{0}]+i[H_{B},\Omega_{0}]+i[H_{2},\Omega_{1}]+...
    \end{aligned}
\end{equation}
We now choose $\Omega_{1}$ such that 
\begin{equation}
    \label{eq:x}
    \begin{aligned} i[H_{1},\Omega_{0}]+i[H_{2},\Omega_{1}]=0,
    \end{aligned}
\end{equation}
and it is obvious that $\Omega_{1}$ is of $O(\frac{1}{\mu})$. In general our decoupling condition will become:
\begin{equation}
    \label{eq:x}
    \begin{aligned} i[H_{1},\Omega_{n}]+i[H_{2},\Omega_{n+1}]=0,
    \end{aligned}
\end{equation}
with $\Omega_{n+1} \sim \frac{1}\mu \Omega_{n}$. Thus, we see that decoupling can be consistently carried out iteratively in a perturbative fashion. 
\subsection{Decoupled Hamiltonian and S-matrix Elements}
\label{subsec:ch22}
Putting together the expansions given above and using the decoupling conditions, it is straightforward to calculate $H_{decoupled}$: 
\begin{equation}
    \label{eq:H(decoup)}
    \begin{aligned} H_{decoupled}=&\bra{0_{H}} H_{1}+H_{2}+H_{B}+\frac{i}{2}[H_{A},\Omega_{0}]+i[H_{B},\Omega_{0}]-\frac{1}{3}[[H_{A},\Omega_{0}],\Omega_{0}]\\
    &-\frac{1}{2}[[H_{B},\Omega_{0}],\Omega_{0}]-\frac{1}{2}[[H_{1},\Omega_{0}],\Omega_{0}]+i[H_{B},\Omega_{1}]+O(\frac{1}{\mu}) \ket{0_{H}}.
    \end{aligned}
\end{equation}
As we pointed out in section~\ref{subsec:ch21}, this $H_{decoupled}$ is not the physical effective Hamiltonian operator at low energies. The path to obtain the effective Hamiltonian will be discussed in the next subsection. 
Here we wish to point out a useful connection of the various terms in Eq.(\ref{eq:H(decoup)}) with corresponding ones in Feynman diagram calculations. Explicit evaluation of the various contributions using mode expansions will be done in section (\ref{sec:Calc}). 

\begin{figure}[tbp]
\centering 
\subfigure[]{\includegraphics[width=.4\textwidth]{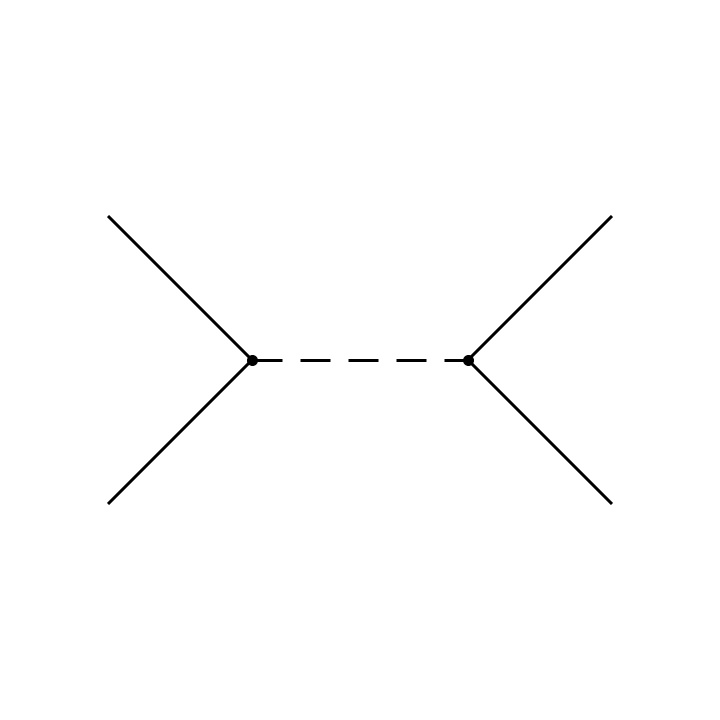}\label{fig:ch221a}}
\hfill
\subfigure[]{\includegraphics[width=.4\textwidth]{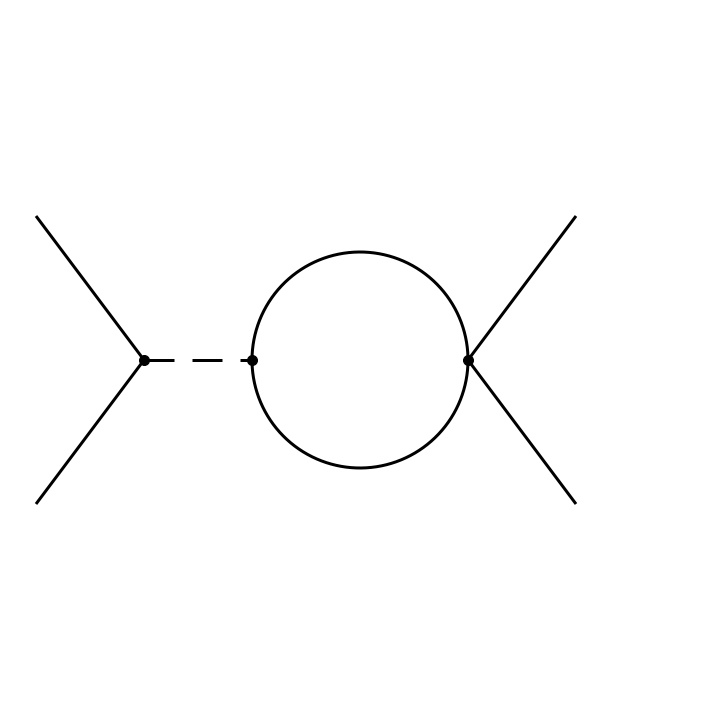}\label{fig:ch221b}}
\caption{\label{fig:i} Two examples of two-to-two scattering process in a scalar field theory with solid lines representing light fields $\Phi_{L}$ and the dashed line representing a heavy particle $\Phi_{H}$. The coupling coefficient at the vertex is $\lambda$. The heavy particle mass M is much larger than the momentum p on the propagator.}
\end{figure}

There are two parts in $H_{decoupled}$. The first part is $\bra{0_{H}}H_{1}\ket{0_{H}}=H_{1}$ which is simply the low energy part in the original Hamiltonian. The second part is due to the contribution of commutators and normal ordering of $H_{B}$ in the expansion. As shown later in the scalar field theory example, each element in the second part can be understood as an S-matrix element in the full theory but expanded in terms of $\frac{1}{\mu}$. For instance, suppose we have the tree level scattering process represented in Figure \ref{fig:ch221a}, the S-matrix element in the full theory including all channels is $-3\lambda^{2}\frac{i}{p^{2}-M^{2}}$, and the corresponding term in $H_{decoupled}$ will be $(-\frac{3\lambda^{2}}{M^{2}}+O(\frac{1}{M^{4}}))\frac{\Phi_{L}^{4}}{4 !}$. This correspondence remains true for the one loop scattering process as well. For example, consider the diagram shown in Figure \ref{fig:ch221b},at order $O(\frac{1}{M^{2}})$. A traditional calculation will give $-\frac{3\lambda_{1}^{2}\lambda_{0}}{16\pi^{2}M^{2}}\left(\frac{1}{\overline{\epsilon}}-\ln\frac{m^{2}}{\mu^{2}}\right)$, and the corresponding contributing terms in $H_{decoupled}$ are a combination of $-\frac{1}{3}[[H_{A},\Omega_{0}],\Omega_{0}]$
    and $\frac{1}{2}[[H_{B},\Omega_{0}],\Omega_{0}]$. We will show in section (\ref{subsubsec:ch322})that this latter contribution is also $-\frac{3\lambda_{1}^{2}\lambda_{0}}{16\pi^{2}M^{2}} \left(\frac{1}{\overline{\epsilon}}-\ln\frac{m^{2}}{\mu^{2}}\right) \frac{\phi^{4}(x)}{4!}$, where the factor 3 appears again because of the summation of contributions from all three channels. This correspondence will continue to hold for all terms in $H_{decoupled}$ as we will see in Section 3. In this sense, we are able to make a term by term correspondence between the traditional Feynman diagram calculations and the contributions in the Hamiltonian formulation of this paper of the terms in Eq.(\ref{eq:H(decoup)}). Due to this connection between the decoupled Hamiltonian and S-matrix elements, Feynman diagrams in the full theory provide useful guidance in organizing practical calculations, as we will see in section \ref{sec:Calc}.

\tikzset{every picture/.style={line width=0.5pt}} 
\begin{figure}
    \centering
\begin{tikzpicture}[x=0.75pt,y=0.75pt,yscale=-1,xscale=1]

\draw   (163,1) -- (297.5,1) -- (297.5,79) -- (163,79) -- cycle ;
\draw   (163,136) -- (297.5,136) -- (297.5,214) -- (163,214) -- cycle ;
\draw    (227.5,79) -- (228.47,136) ;
\draw [shift={(228.5,138)}, rotate = 269.03] [color={rgb, 255:red, 0; green, 0; blue, 0 }  ][line width=0.75]    (10.93,-3.29) .. controls (6.95,-1.4) and (3.31,-0.3) .. (0,0) .. controls (3.31,0.3) and (6.95,1.4) .. (10.93,3.29)   ;
\draw   (394,135) -- (528.5,135) -- (528.5,213) -- (394,213) -- cycle ;
\draw    (297,172) -- (392.5,172) ;
\draw [shift={(394.5,172)}, rotate = 180] [color={rgb, 255:red, 0; green, 0; blue, 0 }  ][line width=0.75]    (10.93,-3.29) .. controls (6.95,-1.4) and (3.31,-0.3) .. (0,0) .. controls (3.31,0.3) and (6.95,1.4) .. (10.93,3.29)   ;
\draw   (163,272) -- (297.5,272) -- (297.5,350) -- (163,350) -- cycle ;
\draw   (281,438) -- (415.5,438) -- (415.5,516) -- (281,516) -- cycle ;
\draw   (394,270) -- (528.5,270) -- (528.5,348) -- (394,348) -- cycle ;
\draw    (229.5,213) -- (230.47,270) ;
\draw [shift={(230.5,272)}, rotate = 269.03] [color={rgb, 255:red, 0; green, 0; blue, 0 }  ][line width=0.75]    (10.93,-3.29) .. controls (6.95,-1.4) and (3.31,-0.3) .. (0,0) .. controls (3.31,0.3) and (6.95,1.4) .. (10.93,3.29)   ;
\draw    (460.5,212) -- (461.47,269) ;
\draw [shift={(461.5,271)}, rotate = 269.03] [color={rgb, 255:red, 0; green, 0; blue, 0 }  ][line width=0.75]    (10.93,-3.29) .. controls (6.95,-1.4) and (3.31,-0.3) .. (0,0) .. controls (3.31,0.3) and (6.95,1.4) .. (10.93,3.29)   ;
\draw    (230.5,392) -- (462.5,391) ;
\draw    (230,349) -- (230.5,392) ;
\draw    (462,348) -- (462.5,391) ;
\draw    (346.5,391.5) -- (348.41,436) ;
\draw [shift={(348.5,438)}, rotate = 267.54] [color={rgb, 255:red, 0; green, 0; blue, 0 }  ][line width=0.75]    (10.93,-3.29) .. controls (6.95,-1.4) and (3.31,-0.3) .. (0,0) .. controls (3.31,0.3) and (6.95,1.4) .. (10.93,3.29)   ;

\draw (185,24) node [anchor=north west][inner sep=0.75pt]   [align=left] {Hamiltonian in\\full theory};
\draw (198,163.4) node [anchor=north west][inner sep=0.75pt]    {$H^{tree}_{decoupled}$};
\draw (85.03,95.16) node [anchor=north west][inner sep=0.75pt]  [rotate=-0.2]  {$\bra{0_{H}} \omega ^{\dagger } H\omega \ket{0_{H}}_{tree}$};
\draw (438,162.4) node [anchor=north west][inner sep=0.75pt]    {$H^{tree}_{eff}$};
\draw (309,145) node [anchor=north west][inner sep=0.75pt]   [align=left] {projection};
\draw (197,297.4) node [anchor=north west][inner sep=0.75pt]    {$H^{oneloop}_{decoupled}$};
\draw (426,295.4) node [anchor=north west][inner sep=0.75pt]    {$H^{oneloop\ eff}_{decoupled}$};
\draw (323,465.4) node [anchor=north west][inner sep=0.75pt]    {$H^{oneloop}_{eff}$};
\draw (310,363) node [anchor=north west][inner sep=0.75pt]   [align=left] {subtraction};
\draw (88.03,229.16) node [anchor=north west][inner sep=0.75pt]  [rotate=-0.2]  {$\bra{0_{H}} \omega ^{\dagger } H\omega \ket{0_{H}}_{one loop}$};
\draw (463.03,229.16) node [anchor=north west][inner sep=0.75pt]  [rotate=-0.2]  {$\bra{0_{H}} \omega ^{'\dagger } H^{tree}_{eff} \omega^{'} \ket{0_{H}}_{one loop}$};
\end{tikzpicture}
\caption{\label{fig:ch231} A flow chart showing how to do matching in our theory and obtain the effective Hamiltonian at one loop order.}
\end{figure}
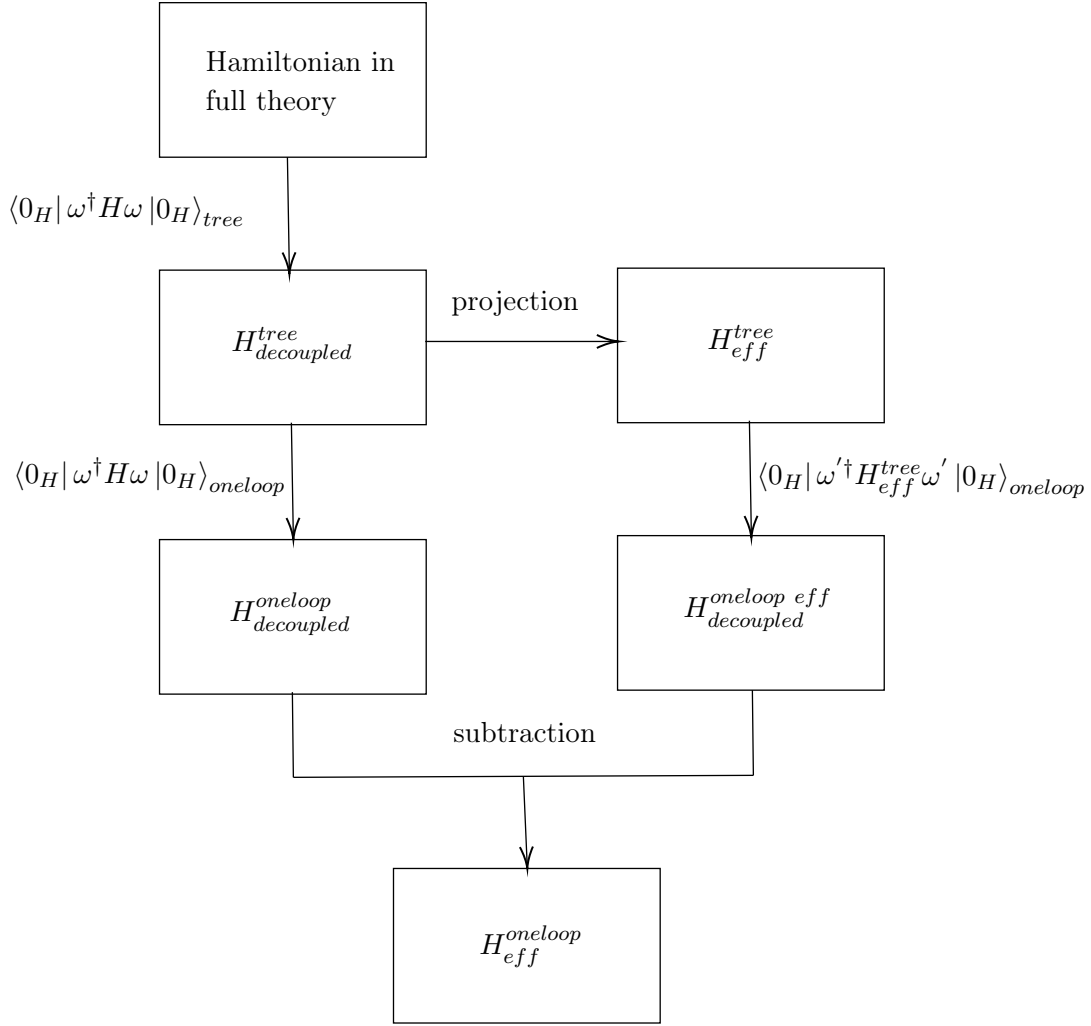

\subsection{Renormalization and Matching}
\label{subsec:ch23}
Since the second part in $H_{decoupled}$ produces terms similar to the  expansion of S-matrix elements, inevitably, there are UV divergences emerging from the loop calculations. Because only low energy modes can appear in $H_{decoupled}$, all the UV divergences should be canceled by the renormalization of $H_{1}$. In this way, we can determine the renormalization Z-factor of light field, light field mass and coupling constants of purely light interactions. It is shown in section \ref{subsec:ch33} that our results obtained from $H_{decoupled}$ indeed agree with those obtained using traditional renormalization in the Lagrangian framework.\\
\indent The effective Hamiltonian is obtained by matching order by order as we show in Figure \ref{fig:ch231}. First we decouple the full theory at tree level by doing tree level calculation using the unitary transformation $H_{decoupled}^{tree}=\bra{0_{H}} \omega^{\dagger} H \omega \ket{0_{H}}_{tree}$ and match it onto the low energy regime to get the tree level effective Hamiltonian. This step is simple, we can just denote $H_{eff}^{tree}=H_{decoupled}^{tree}$. Then we obtain the decoupled Hamiltonian at one loop in both full and effective theories by calculating $\bra{0_{H}} \omega^{\dagger} H \omega \ket{0_{H}}_{one loop}$ and $\bra{0_{H}} \omega^{'\,\dagger} H_{eff}^{tree} \omega^{'} \ket{0_{H}}_{one loop}$ at one loop order, where $\omega^{'}$ is the decoupling unitary transformation for the effective theory. After renormalizing both full and effective theories, we will get decoupled Hamiltonians $H_{decoupled}^{one loop}$ and $H_{decoupled}^{one loop,eff}$ which include interaction terms that account for the one loop corrections. Similar to the traditional EFT, we can do the matching by subtracting $H_{decoupled}^{one loop,eff}$ from $H_{decoupled}^{one loop}$ to get new interaction terms to be added in the effective Hamiltonian at one loop order which now does not contain any large logarithmic contributions. We can proceed to higher orders iteratively in this fashion.

\section{Decoupling and Renormalization of A Scalar Field Theory}
\label{sec:Calc}
In this section we will work in the weak coupling regime of a scalar field theory with both heavy and light fields. Because we will be discussing renormalization in the Hamiltonian framework, we will consider mode expansions at a fixed time or effectively, we will be working in the Schrodinger picture. Also, for the purpose of decoupling, we set the cut-off energy scale $\mu$ to be the heavy mass M.
\subsection{Preliminaries}
\label{subsec:ch31}
Our subsequent analysis will apply to a scalar field theory with heavy and light fields ($\Phi_H$ and $\Phi_L$ respectively) with dynamics given by the following Hamiltonian:\\
\begin{equation}
\label{eq:x}
    \begin{aligned} H=& \int d^{3} x \left(\frac{1}{2}\left(\Pi_{L}^{2}(x)+(\nabla \Phi_{L}(x))^{2}+m^{2}\Phi_{L}^{2}(x)\right)+\frac{1}{2}\left(\Pi_{H}^{2}(x)+(\nabla \Phi_{H}(x))^{2}+M^{2}\Phi_{H}^{2}(x)\right)\right.\\
    &\left.+\frac{\lambda_{0}}{4 !} \Phi_{L}^{4}(x)+\frac{\lambda_{1}}{2} \Phi_{H}(x) \Phi_{L}^{2}(x)+\frac{\lambda_{2}}{4} \Phi_{L}^{2}(x) \Phi_{H}^{2}(x)+\frac{\lambda_{3}}{4 !} \Phi_{H}^{4}(x) 
    \right).
    \end{aligned}
\end{equation}
\indent These fields have the usual mode expansions, however, we will need to consider light fields carefully. This is because the light fields contain two parts, one is the low frequency mode $\phi(x)$ and the other is the high frequency mode $\chi (x)$. In order to correctly project onto the low energy subspace, we want only low frequency fields $\phi(x)$ to appear in external lines. This can be taken into account in the usual expansion of all the fields (in Schrodinger picture) in the following manner:\\
\begin{subequations}\label{eq:y}
\begin{align}
\label{eq:y:1}
\phi(x)=\sum_{p<M} \frac{1}{\sqrt{2 \epsilon_{p}}}\left(b_{p} e^{i \boldsymbol p \boldsymbol x} +b_{p}^{\dagger} e^{-i \boldsymbol p \boldsymbol x}\right),
\\
\label{eq:y:2}
\chi(x)=\sum_{M<p} \frac{1}{\sqrt{2 \epsilon_{p}}}\left(b_{p} e^{i \boldsymbol p \boldsymbol x} +b_{p}^{\dagger} e^{-i \boldsymbol p \boldsymbol x}\right),
\\
\label{eq:4:3}
\Phi_{L}(x)=\sum_{p} \frac{1}{\sqrt{2 \epsilon_{p}}}\left(b_{p} e^{i \boldsymbol p \boldsymbol x} +b_{p}^{\dagger} e^{-i \boldsymbol p \boldsymbol x}\right),
\\
\label{eq:y:4}
\Phi_{H}(x)=\sum_{k} \frac{1}{\sqrt{2 \omega_{k}}}\left(a_{k} e^{i \boldsymbol k \boldsymbol x} +a_{k}^{\dagger} e^{-i \boldsymbol k \boldsymbol x}\right).
\end{align}
\end{subequations}
In the above, and for the rest of this paper, we adopt the notation that $\omega_k$ denotes the energy of the heavy particle and $\epsilon_p$ that of the light one. Thus, for example, $\omega_k=\sqrt{\vec{k}^2+M^2}$ and $\epsilon_p=\sqrt{\vec{p}^2+m^2}$.
From the expansion, we see that the $\phi$ and the $\chi$ fields are orthogonal, i.e., $\int d^{3}x\, \phi(x) \chi(x)=0$. In the following we will not use the mode expansion of $\phi(x)$ . Finally, as a notational convenience, in going from the discrete momentum sum to the continuum we will use $\sum_{k} \rightarrow \int \frac{d^3k}{(2\pi)^3}$,and omit all factors of the volume $V$ since these will eventually cancel out.\\
\indent As discussed earlier, we want to split the total Hamiltonian into four parts: $H_{1}$ contains only low frequency modes of light particles; $H_{2}$ contains the free parts of both heavy particles and high frequency modes of light particles; $H_{A}$ contains only creation or annihilation operators, e.g., $aab$, $a^{\dagger}a^{\dagger}b^{\dagger}$, etc; and $H_{B}$ contains combinations of creation and annihilation operators, e.g., $b^{\dagger}a\phi(x)$, $a^{\dagger}abb$, etc. Thus,
\begin{align}
H = H_1 + H_2 + H_A + H_B.
\end{align}
For our case,
\begin{subequations}\label{eq:y}
\begin{align}
\label{eq:y:1}
H_{1}&=\int d^{3} x \left(
    \frac{1}{2}\left(\Pi^{2}+(\nabla \phi)^{2}+m^{2}\phi^{2}\right)+\frac{\lambda_{0}}{4 !} \phi^{4}(x)\right),\\
\label{eq:y:2}
H_{2}&=\sum_{k} \omega_{k}a_{k}^{\dagger}a_{k}+\sum_{M<p}\epsilon_{p}b_{p}^{\dagger}b_{p}.
\end{align}
\end{subequations}
However, $H_{A}$ and $H_{B}$ are rather involved and will not be explicitly displayed here. As we proceed with the calculation, we will simply pick out relevant terms by analyzing the coupling coefficient and the number of low energy light particles in external legs.  \\
\indent We argued earlier that $H_{decoupled}=\bra{0_{H}} \omega^{\dagger} H \omega \ket{0_{H}}$, where $\omega$ denotes a series of unitary transformations, $\omega=\omega_{0} \omega_{1} \dots \omega_{n} \dots$ Our calculation will be limited to the first loop order and for this purpose, we only need the first two terms in the unitary transformations:\\
\begin{equation}\label{eq:x}
    \begin{aligned} \omega^{\dagger} H \omega = e^{-i \Omega_{1}} e^{-i \Omega_{0}}\left(H_{1}+H_{2}+H_{A}+H_{B}\right) e^{i \Omega_{0}} e^{i \Omega_{1}}.
    \end{aligned}
\end{equation}
\indent The right hand side of the above equation can be simplified to\\ 
\begin{equation}\label{Hdecoup}
    \begin{aligned} &H_{1}+H_{2}+H_{B}+\frac{i}{2}[H_{A},\Omega_{0}]+i[H_{B},\Omega_{0}]-\frac{1}{3}[[H_{A},\Omega_{0}],\Omega_{0}]\\
    &-\frac{1}{2}[[H_{B},\Omega_{0}],\Omega_{0}]-\frac{1}{2}[[H_{1},\Omega_{0}],\Omega_{0}]+i[H_{B},\Omega_{1}]+O(\frac{1}{M}),
    \end{aligned}
\end{equation}
where we have set the cut-off energy scale to the heavy mass M and used the condition $i\left[H_{2}, \Omega_{0}\right] + H_{A} = 0$, $i\left[H_{1}, \Omega_{0}\right] + i\left[H_{2}, \Omega_{1}\right] = 0$.

In the next two sections we will study decoupling and renormalization in this scalar field theory by calculating the decoupled Hamiltonian up to order $O(\frac{1}{M^{2}})$ at one loop level for the two and four point functions. The techniques involved in the calculation are different from the traditional Feynman diagram methods, however, as we have noticed earlier, Feynman diagrams can provide a good indication of which term in Eq.(\ref{Hdecoup}) contributes to the process of interest. Thus in the following, even though we are not using the usual Feynman-Dyson perturbative expansion, we will still refer to the corresponding Feynman diagrams in guiding the choice of the relevant terms in Eq.(\ref{Hdecoup}).
\subsection{Decoupling}
In this section we will explicitly calculate the decoupled Hamiltonian from Eq.(\ref{Hdecoup}), for the light particle two and four point functions at the one loop level. To simplify the notations we will take it as understood that all operator expressions must be sandwiched between the high energy vacuum state to obtain the decoupled Hamiltonian from Eq.(\ref{Hdecoup}). 
\label{subsec:ch32}
\subsubsection{Two Point Function of Light Fields}
\label{subsubsec:ch321}
\begin{figure}[tbp]
\centering 
\subfigure[]{\includegraphics[width=.2\textwidth]{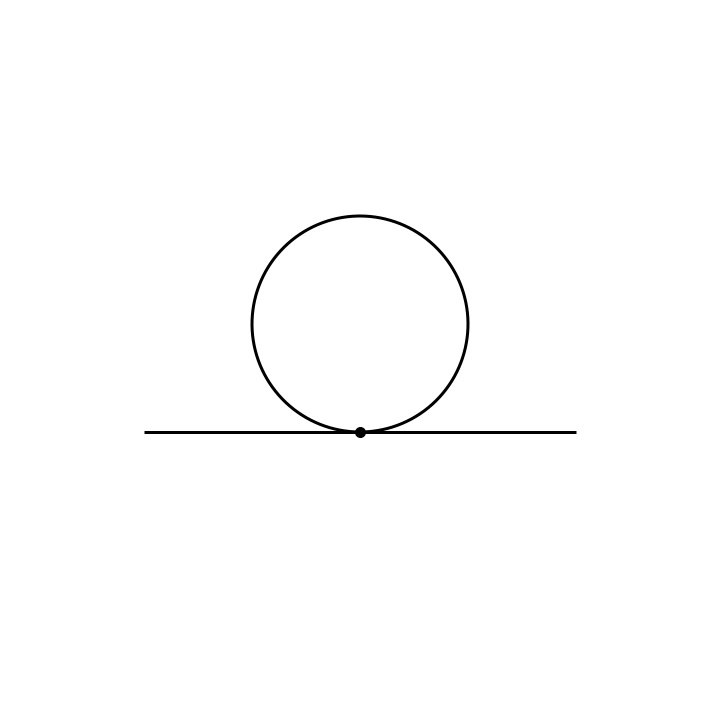}\label{fig:ch31a}}
\hfill
\subfigure[]{\includegraphics[width=.2\textwidth]{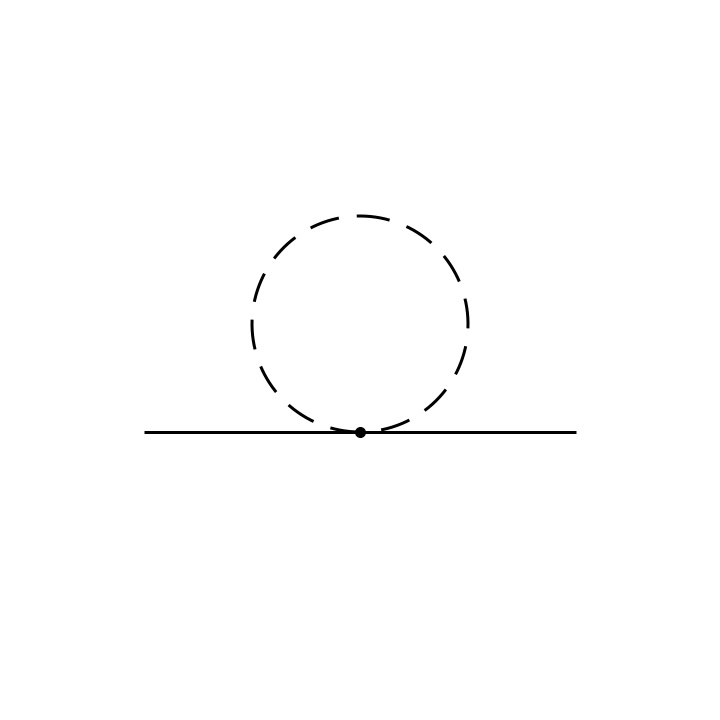}\label{fig:ch31d}}
\hfill
\subfigure[]{\includegraphics[width=.2\textwidth]{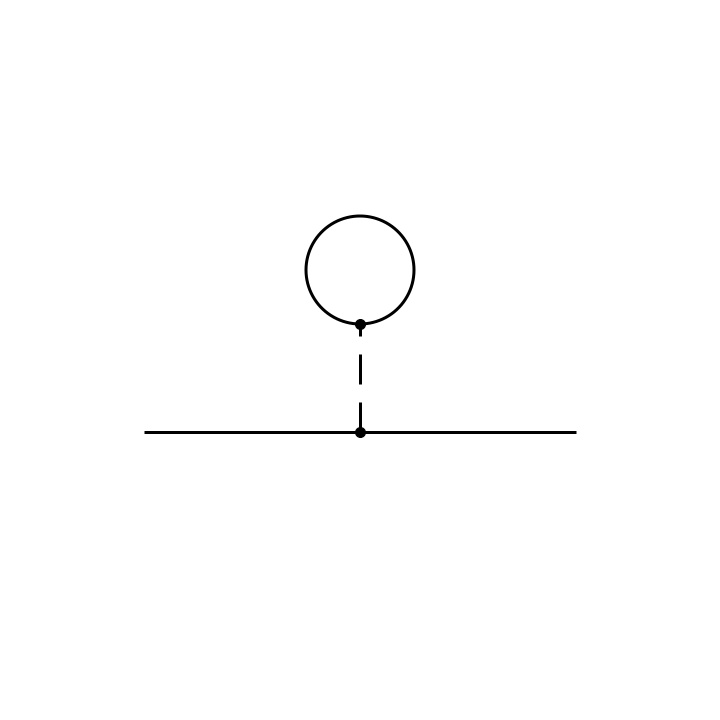}\label{fig:ch31c}}
\hfill
\subfigure[]{\includegraphics[width=.2\textwidth]{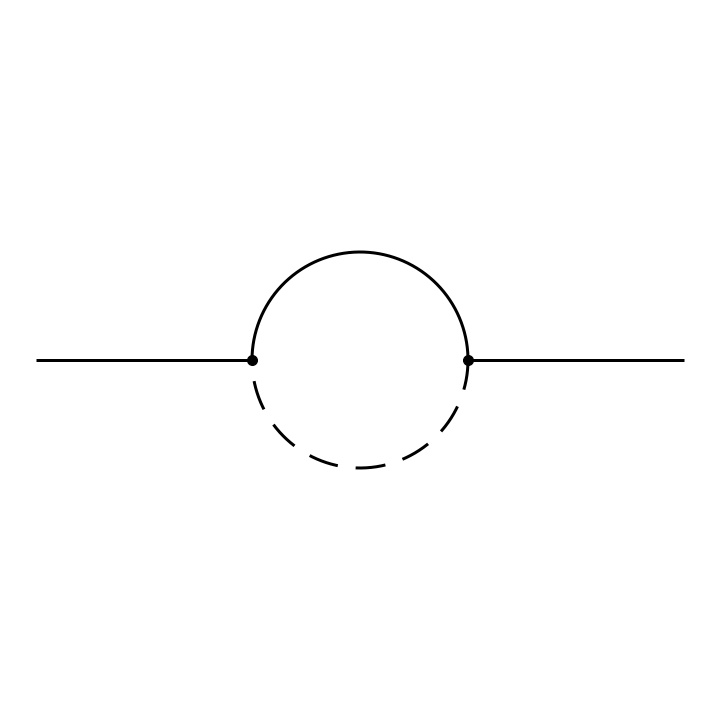}\label{fig:ch31b}}
\caption{\label{fig:i} One loop two point functions. Dashed lines represent heavy fields and solid lines represent light fields.}
\end{figure}
There are four contributions to the two point function at one loop order which we choose to specify by ordinary Feynman diagrams. Let's consider the Figure \ref{fig:ch31a} which comes from the $\frac{\lambda_{0}}{4!}\Phi_{L}^{4}(x)$ term in the total Hamiltonian and make the mode expansion for the $\phi^{2}\chi^{2}$ piece:
\begin{equation}\label{eq:x}
    \begin{aligned} 6\frac{\lambda_{0}}{4 !} \int d^{3}x\, \sum_{M<k}\sum_{M<p} \frac{1}{\sqrt{2\epsilon_{k}}}\frac{1}{\sqrt{2\epsilon_{p}}} e^{i\left(\boldsymbol k - \boldsymbol p \right) \boldsymbol x} b_{k}b_{p}^{\dagger} \phi^{2}(x).
    \end{aligned}
\end{equation}
Normal ordering this term gives\\
\begin{equation}\label{eq:x}
    \begin{aligned} 6\frac{\lambda_{0}}{4 !} \int d^{3}x\, \sum_{M<k}\sum_{M<p} \frac{1}{\sqrt{2\epsilon_{k}}}\frac{1}{\sqrt{2\epsilon_{p}}} e^{i\left(\boldsymbol k - \boldsymbol p \right) \boldsymbol x} \left(\left[b_{k},b_{p}^{\dagger} \right]+b_{p}^{\dagger}b_{k}\right) \phi^{2}(x).
    \end{aligned}
\end{equation}
Keeping the commutator piece since only this survives upon sandwiching the above expression between high energy vacuum states, we get\\
\begin{equation}\label{eq:x}
    \begin{aligned}\frac{\lambda_{0}}{8} \int d^{3}x \int \frac{d^{3}k}{\left(2 \pi\right)^{3}} \, \frac{\phi^{2}(x)}{\sqrt{k^{2}+m^{2}}}-\frac{\lambda_{0}}{8} \int d^{3}x \int_{k<M} \frac{d^{3}k}{(2\pi)^{3}} \, \frac{\phi^{2}(x)}{\sqrt{k^{2}+m^{2}}}.
    \end{aligned}
\end{equation}
We use dimensional regularization in $d=3-2\epsilon$ dimensions, where usual UV divergence appears as a pole in Hamiltonian framework at $d=3$. The result is\\
\begin{equation}\label{eq:x}
    \begin{aligned} &-\int d^{3}x \, \frac{\phi^{2}(x)}{2} \frac{\lambda_{0} m^{2}}{32 \pi^{2}}\left(\frac{1}{\overline{\epsilon}}-\ln{\frac{m^{2}}{\mu^{2}}}+1\right)+\frac{\lambda_{0}}{4}C\int d^{3}x \, \frac{\phi^{2}(x)}{2},\\
    & \frac{1}{\overline{\epsilon}}=\frac{1}{\epsilon}-\gamma+\ln{4\pi},\\
    &C=-\int_{k<M}\frac{d^{3}k}{(2\pi)^{3}}\,\frac{1}{\sqrt{k^{2}+m^{2}}}.
    \end{aligned}
\end{equation}
The term proportional to C arises from the restriction imposed on the momentum of the high frequency part of the light field, $\chi(x)$, namely $p>M$. In other words, in order to use dimensional regularization we extend the momentum integral to the full range and subtract the infrared region, whose contribution is $C$. This procedure will be followed for all such integrals and we will show later that finite infrared contributions like $C$ will cancel out when we do the matching to construct the effective Hamiltonian.\\
\indent Similarly, the Figure \ref{fig:ch31d} arises from the term\\
\begin{equation}\label{eq:x}
    \begin{aligned} \frac{\lambda_{2}}{4} \int d^{3}x\, \sum_{k}\sum_{p} \frac{1}{\sqrt{2\omega_{k}}}\frac{1}{\sqrt{2\omega_{p}}} e^{i\left(\boldsymbol k - \boldsymbol p \right) \boldsymbol x} a_{k}a_{p}^{\dagger} \phi^{2}(x).
    \end{aligned}
\end{equation}
Normal ordering gives,\\
\begin{equation}\label{eq:x}
    \begin{aligned} \frac{\lambda_{2}}{4} \int d^{3}x\, \sum_{k}\sum_{p} \frac{1}{\sqrt{2\omega_{k}}}\frac{1}{\sqrt{2\omega_{p}}} e^{i\left(\boldsymbol k - \boldsymbol p \right) \boldsymbol x} \left(\left[a_{k},a_{p}^{\dagger} \right]+a_{p}^{\dagger}a_{k}\right) \phi^{2}(x).
    \end{aligned}
\end{equation}
Using dimensional regularization, we can get as the contribution to the decoupled Hamiltonian, \\
\begin{equation}\label{eq:x}
    \begin{aligned} -\int d^{3}x \, \frac{\phi^{2}(x)}{2} \frac{\lambda_{2} M^{2}}{32 \pi^{2}}\left(\frac{1}{\overline{\epsilon}}-\ln{\frac{M^{2}}{\mu^{2}}}+1\right).
    \end{aligned}
\end{equation}
\indent Figure \ref{fig:ch31c} is proportional to $\lambda_{1}^{2}$ and $\phi^{2}$. Since it is second order in coupling constant, it must come from the term $\frac{i}{2}\left[H_{A}, \Omega_{0}\right]$ in Eq.(\ref{Hdecoup}). Let us denote the corresponding $H_{A}$ in this case by $H_{A}^{2,1}$ and consider the mode expansion of $\frac{\lambda_{1}}{2} \Phi_{H}\Phi_{L}^{2}$\\
\begin{equation}\label{eq:3.14}
    \begin{aligned} \frac{\lambda_{1}}{2} \Phi_{H} \Phi_{L}^{2}&= \frac{\lambda_{1}}{2} \int d^{3}x \, \sum_{k} \sum_{M<p} \sum_{M<q} \frac{1}{2^{\frac{3}{2}}\sqrt{\omega_{k}\epsilon_{p}\epsilon_{q}}} \left(a_{k} e^{i \boldsymbol k \boldsymbol x} + a_{k}^{\dagger} e^{-i \boldsymbol k \boldsymbol x}\right)\\ 
    &\left(b_{p} e^{i \boldsymbol p \boldsymbol x} + b_{p}^{\dagger} e^{-i \boldsymbol p \boldsymbol x} \right) \left(b_{q} e^{i \boldsymbol q \boldsymbol x} + b_{q}^{\dagger} e^{-i \boldsymbol q \boldsymbol x} \right)+\frac{\lambda_{1}}{2} \int d^{3}x \, \sum_{k} \frac{1}{\sqrt{2\omega_{k}}}\\ &
    \phi^{2}(x)\left(a_{k} e^{i \boldsymbol k \boldsymbol x} + a_{k}^{\dagger} e^{-i \boldsymbol k \boldsymbol x}\right)+...
    \end{aligned}
\end{equation}
where the dots represent terms proportional to $\phi(x)$.\\
\indent This expansion has a piece proportional to $\frac{\lambda_{1}}{2} \int d^{3}x \, \sum_{k} \frac{1}{\sqrt{2\omega_{k}}} \left(a_{k} e^{i \boldsymbol k \boldsymbol x} + a_{k}^{\dagger} e^{-i \boldsymbol k \boldsymbol x}\right) \phi^{2}(x)$ which contributes to $H_{A}^{2,1}$. In addition, there's another term, which is in the form of $H_{B}$:\\
\begin{equation}\label{eq:x}
    \begin{aligned} &\frac{\lambda_{1}}{2} \int d^{3}x \, \sum_{k} \sum_{M<p} \sum_{M<q} \frac{1}{2^{\frac{3}{2}}\sqrt{\omega_{k}\epsilon_{p}\epsilon_{q}}} \left(a_{k} e^{i \boldsymbol k \boldsymbol x} + a_{k}^{\dagger} e^{-i \boldsymbol k \boldsymbol x}\right) b_{p} e^{i \boldsymbol p x} b_{q}^{\dagger} e^{-i \boldsymbol q \boldsymbol x}.\\
    \end{aligned}
\end{equation}
Normal ordering this gives,
\begin{equation}
    \begin{aligned} \frac{\lambda_{1}}{2} \int d^{3}x \, \sum_{k} \sum_{M<p} \sum_{M<q} \frac{1}{2^{\frac{3}{2}}\sqrt{\omega_{k}\epsilon_{p}\epsilon_{q}}} \left(a_{k} e^{i \boldsymbol k \boldsymbol x} + a_{k}^{\dagger} e^{-i \boldsymbol k \boldsymbol x}\right) e^{i \left(\boldsymbol k -\boldsymbol p\right) \boldsymbol x} \left(\left[b_{p},b_{q}^{\dagger}\right]+b_{q}^{\dagger} b_{p}\right).
    \end{aligned}
\end{equation}
Evaluating this we get the net contribution for $H_{A}^{2,1}$ to be  
\begin{equation}\label{eq:x}
    \begin{aligned} H_{A}^{2,1} = \int d^{3}x \, \frac{\lambda_{1}}{2} \sum_{k} \frac{1}{\sqrt{2\omega_{k}}} \left(a_{k} e^{i \boldsymbol k \boldsymbol x} + a_{k}^{\dagger} e^{-i \boldsymbol k \boldsymbol x}\right) \left(-\frac{m^{2}}{16 \pi^{2}} \left(\frac{1}{\overline{\epsilon}}-\ln{\frac{m^{2}}{\mu^{2}}}+1\right)+\frac{C}{2} +\phi^{2}(x)\right),
    \end{aligned}
\end{equation}
where again $C=-\int_{k<M}\frac{d^{3}k}{(2\pi)^{3}}\,\frac{1}{\sqrt{k^{2}+m^{2}}}$ as before.
We next evaluate $\Omega_{0}$ for this case by which we denote by $\Omega_{0}^{2,1}$. This is obtained from from the condition  $i\left[H_{2},\Omega_{0}^{2,1}\right]+H_{A}^{2,1}=0$ which gives,\\
\begin{equation}\label{eq:x}
    \begin{aligned} \Omega_{0}^{2,1} = \int d^{3}y \, \frac{\lambda_{1}}{2} \sum_{p} \frac{-i}{\sqrt{2\omega_{p}}\omega_{p}} \left(a_{p} e^{i \boldsymbol p \boldsymbol y} - a_{p}^{\dagger} e^{-i \boldsymbol p \boldsymbol y}\right) \left(-\frac{m^{2}}{16 \pi^{2}} \left(\frac{1}{\overline{\epsilon}}-\ln{\frac{m^{2}}{\mu^{2}}}+1\right)+\frac{C}{2} +\phi^{2}(y)\right).
    \end{aligned}
\end{equation}
From this we obtain,
\begin{equation}\label{eq:x}
    \begin{aligned} \frac{i}{2}\left[H_{A}^{2,1},\Omega_{0}^{2,1}\right]&= -\frac{i}{2} \int \int d^{3}x d^{3}y \, \frac{\lambda_{1}^{2}}{4} \sum_{k} \sum_{p} \frac{-i}{2\sqrt{\omega_{k} \omega_{p}}\omega_{p}} \left[a_{k},-a_{p}^{\dagger}\right] e^{i \boldsymbol k \boldsymbol x} e^{-i \boldsymbol p \boldsymbol y} \\& \left(\frac{m^{2}}{16 \pi^{2}} \left(\frac{1}{\overline{\epsilon}}-\ln{\frac{m^{2}}{\mu^{2}}}+1\right)-\frac{C}{2}\right) \phi^{2}(y) + \dots \\
    &= \int \int d^{3}x d^{3}y ~\frac{\lambda_{1}^{2}}{16}\int \frac{d^{3}k}{{(2\pi)}^3} \frac{1}{\omega_{k}^{2}} e^{i \boldsymbol k \left(\boldsymbol x - \boldsymbol y\right)}\left(\frac{m^{2}}{16 \pi^{2}} \left(\frac{1}{\overline{\epsilon}}-\ln{\frac{m^{2}}{\mu^{2}}}+1\right)-\frac{C}{2}\right) \phi^{2}(y)+\dots,\\
    \end{aligned}
\end{equation}
where the dots denote three other similar terms arising from the commutator.
 The integral over $x$ forces the momentum k to zero, and $\omega_{k}^{2} \rightarrow M^{2}$. Note that in terms of the associated Feynman diagram, $k$ is the (zero) external momenta flowing into the vacuum bubble of the tadpole. This association with the external momentum flow in Feynman diagrams is a general feature of the terms in $H_{decoupled}$, as we will see. Including all these contributions we get for this contribution to the decoupled Hamiltonian,
\begin{equation}\label{eq:x}
    \begin{aligned} \frac{i}{2}\left[H_{A}^{2,1},\Omega_{0}^{2,1}\right]&= \int d^{3}y \frac{4 \lambda_{1}^{2}}{16M^2} \left(\frac{m^{2}}{16 \pi^{2}} \left(\frac{1}{\overline{\epsilon}}-\ln{\frac{m^{2}}{\mu^{2}}}+1\right)-\frac{C}{2}\right) \phi^{2}(y)\\
    &=  \int d^{3}x \, \frac{\lambda_{1}^{2}m^{2}}{32 \pi^{2} M^{2}} \left(\frac{1}{\overline{\epsilon}}-\ln{\frac{m^{2}}{\mu^{2}}}+1\right) \frac{\phi^2(x)}{2}-\frac{\lambda_{1}^{2}C}{4M^{2}}\int d^{3}x\, \frac{\phi^{2}(x)}{2}.
    \end{aligned}
\end{equation}
\indent Next, consider Figure \ref{fig:ch31b} which is the last contribution to the two point function at one loop order. Figure \ref{fig:ch31b} has both light particle and heavy particle propagators. It also arises from the expansion of $\frac{\lambda_{1}}{2}\Phi_{H}\Phi_{L}^{2}$, which is proportional to $\phi$, and is not explicitly listed in Eq.(\ref{eq:3.14}). This contribution is\\
\begin{equation}\label{eq:x}
    \begin{aligned} H_{A}^{2,2} = 2 \int d^{3}x \,  \sum_{k} \sum_{M<p} \frac{\lambda_{1}}{2} \frac{\phi(x)}{2\sqrt{\omega_{k}\epsilon_{p}}} \left(a_{k} e^{i \boldsymbol k \boldsymbol x} b_{p} e^{i \boldsymbol p \boldsymbol x} + h.c. \right).
    \end{aligned}
\end{equation}
Similarly, from $i\left[H_{2},\Omega_{0}^{2,2}\right]+H_{A}^{2,2}=0$, we can get the $\Omega_{0}^{2,2}$ that corresponds to $H_{A}^{2,2}$,\\
\begin{equation}\label{eq:x}
    \begin{aligned} \Omega_{0}^{2,2} =2 \int d^{3}y \,  \sum_{q} \sum_{M<r} \frac{\lambda_{1}}{2} \left( \frac{-i \phi(y)}{2\sqrt{\omega_{q}\epsilon_{r}} \left(\omega_{q}+\epsilon_{r}\right)} a_{q} e^{i \boldsymbol q \boldsymbol y} b_{r} e^{i \boldsymbol r \boldsymbol y} + h.c. \right).
    \end{aligned}
\end{equation}
Then, we need to calculate $\frac{i}{2}\left[H_{A}^{2,2},\Omega_{0}^{2,2}\right]$. After a simple calculation, and normal ordering we get:\\
\begin{equation}\label{eq:3.23}
    \begin{aligned} \frac{i}{2}\left[H_{A}^{2,2},\Omega_{0}^{2,2}\right]&= - \int \int d^{3}x d^{3}y \, \sum_{k} \sum_{M<p} \lambda_{1}^{2} \frac{ \phi(x) \phi(y)}{4\omega_{k}\epsilon_{p}\left(\omega_{k}+\epsilon_{p}\right)} e^{i \left(\boldsymbol k + \boldsymbol p\right) \left(\boldsymbol x - \boldsymbol y\right)},\\
    &= - \int \int d^{3}x d^{3}y \, \sum_{k} \sum_{M<p} \lambda_{1}^{2} \frac{ \phi(x) \phi(y)e^{i \left(\boldsymbol k + \boldsymbol p\right) \left(\boldsymbol x - \boldsymbol y\right)}}{4\sqrt{k^{2}+M^{2}}\sqrt{p^{2}+m^{2}}\left(\sqrt{k^{2}+M^{2}}+\sqrt{p^{2}+m^{2}}\right)}.
    \end{aligned}
\end{equation}
This integral is evaluated in Appendix \ref{sec:app31}. The net contribution from Figure \ref{fig:ch31b} to the decoupled Hamiltonian is then found to be\\
\begin{equation}\label{eq:3.24}
    \begin{aligned} 
    &\frac{m^{2}\lambda_{1}^{2}}{16\pi^{2}M^{2}}\left(\frac{1}{\overline{\epsilon}}-\ln{\frac{m^{2}}{\mu^{2}}}+1\right)\int d^{3}x\, \frac{\phi^{2}(x)}{2}-\frac{\lambda_{1}^{2}C}{2M^{2}}\int d^{3}x\,\frac{\phi^{2}(x)}{2}\\
    &-\frac{\lambda_{1}^{2}}{16\pi^{2}}\left(\frac{1}{\overline{\epsilon}}-\ln{\frac{M^{2}}{\mu^{2}}}+1\right)\left(1+\frac{m^{2}}{M^{2}}\right)\int d^{3}x \, \frac{\phi^{2}(x)}{2} +\frac{\lambda_{1}^{2}}{32\pi^{2}M^{2}}\int d^{3}x\,\frac{1}{2} (\nabla \phi(x))^{2}.
    \end{aligned}
\end{equation}
Note that there is a finite kinetic energy correction term in Eq.(\ref{eq:3.24}), which implies that there should as well be a momentum term $\frac{\lambda_{1}^{2}}{32\pi^{2}M^{2}}\int d^{3}x\,\frac{1}{2} \Pi^{2}(x)$ with the same coefficient. In fact, this term can also be calculated, based on Figure \ref{fig:ch31b}. However, it arises from the commutator $\bra{0_{H}}-\frac{1}{2}[[H_{1},\Omega_{0}],\Omega_{1}]\ket{0_{H}}$ in the higher order expansion of $H_{decoupled}$, which we have omitted in Eq.(\ref{Hdecoup}). The exact calculation of this term is given in Appendix \ref{sec:app34} and it confirms the expectations above.
Putting contributions from all the diagrams together and taking $\mu \approx M$ to get rid of log terms $\ln{\frac{M^{2}}{\mu^{2}}}$, we have the net result for two point functions in $H_{decoupled}$ as
\begin{equation}
    \begin{aligned}\label{eq:3.25}
    &-\frac{\lambda_{0}m^{2}}{32\pi^{2}}\left(\frac{1}{\overline{\epsilon}}-\ln{\frac{m^{2}}{M^{2}}}+1\right)\int d^{3}x\, \frac{\phi^{2}(x)}{2}-\frac{\lambda_{2}M^{2}}{32\pi^{2}}\left(\frac{1}{\overline{\epsilon}}+1\right)\int d^{3}x\, \frac{\phi^{2}(x)}{2}\\
    &+\frac{3\lambda_{1}^{2}m^{2}}{32\pi^{2}M^{2}}\left(\frac{1}{\overline{\epsilon}}-\ln{\frac{m^{2}}{M^{2}}}+1\right)\int d^{3}x\, \frac{\phi^{2}(x)}{2}-\frac{\lambda_{1}^{2}}{16\pi^{2}}\left(\frac{1}{\overline{\epsilon}}+1\right)\left(1+\frac{m^{2}}{M^{2}}\right)\int d^{3}x\, \frac{\phi^{2}(x)}{2}\\
    &+\frac{\lambda_{1}^{2}}{32\pi^{2}M^{2}}\int d^{3}x\,\frac{1}{2} (\Pi^{2}(x)+(\nabla \phi(x))^{2})+\left(\frac{\lambda_{0}}{4}-\frac{3\lambda_{1}^{2}}{2M^{2}}\right)C\int d^{3}x\,\frac{\phi^{2}(x)}{2}.
    \end{aligned}
\end{equation}
\indent There are several finite contributions proportional to $C=-\int_{k<M}\frac{d^{3}k}{(2\pi)^{3}}\,\frac{1}{\sqrt{k^{2}+m^{2}}}$, which will not affect the renormalization of $H_{1}$ as we will show in section \ref{subsec:ch33}. In fact, we will show in section \ref{sec:EFT},that they will only appear in $H_{decoupled}$ and not in the physical effective Hamiltonian $H_{eff}$ which is constructed by a matching process. Another way to think about the effects of these finite terms is that whenever we have a light field in the loop, our calculation will produce these finite terms along with (in a linear way) the "troublesome" large log terms $\ln{\frac{m^{2}}{M^{2}}}$. Therefore, as long as the large log terms can be canceled during matching, we can convince ourselves that these extra finite terms will also be canceled and therefore not appear in $H_{eff}$. All this is essentially a reflection of the fact that the infrared structure of the full and effective theory are the same.
\subsubsection{Four Point Function of Light Fields }
\label{subsubsec:ch322}
\begin{figure}[t]
    \centering
    \includegraphics[width=.3\textwidth]{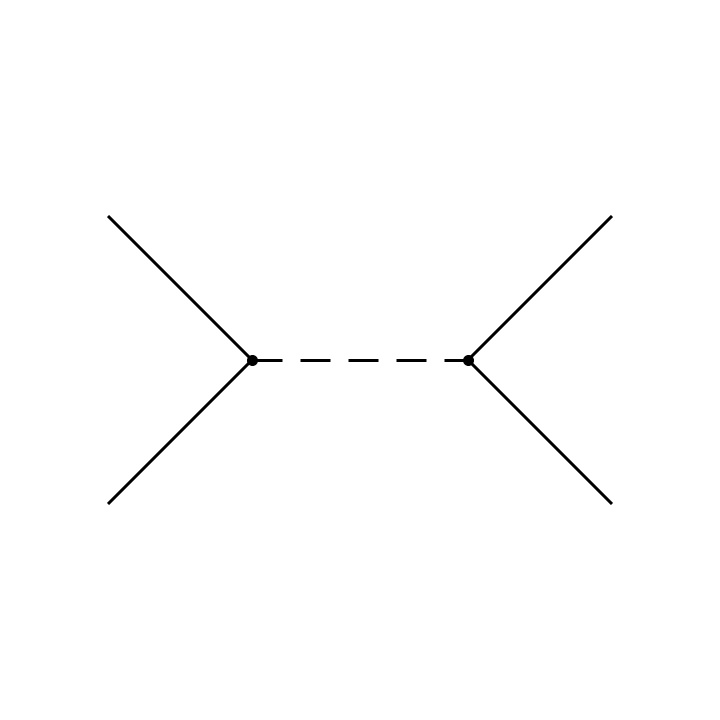}
    \caption{\label{fig:4pt} Four point function at tree level.}
\end{figure}
We will begin with the tree level contribution at order $1/M^2$. In the calculation of Figure \ref{fig:ch31c} in section \ref{subsubsec:ch321}, there is one more term from the commutator $\frac{i}{2}\left[H_{A}^{2,1},\Omega_{0}^{2,1}\right]$ left unexplored that contributes to the four point function at tree level shown in Figure \ref{fig:4pt}:
\begin{equation}
    \begin{aligned} \frac{i}{2}\left[H_{A}^{2,1},\Omega_{0}^{2,1}\right]_{tree}&=i\int d^{3}x \int d^{3} y \,\frac{\lambda_{1}^{2}}{4} \sum_{k} \sum_{p} \frac{-i}{2\sqrt{\omega_{k}\omega_{p}}\omega_{p}}\phi^{2}(x)\phi^{2}(y)[a_{k},-a_{p}^{\dagger}]e^{i\boldsymbol{kx}-i\boldsymbol{py}},\\
    &=-\int d^{3}x\int d^{3}y \int \frac{d^{3}k}{(2\pi)^{3}}\frac{\lambda_{1}^{2}}{8(k^{2}+M^{2})}\phi^{2}(x)\phi^{2}(y)e^{i\boldsymbol{k}(\boldsymbol{x}-\boldsymbol{y})},\\
    &\approx -\frac{3\lambda_{1}^{2}}{M^{2}}\int d^{3}x\,\frac{\phi^{4}(x)}{4 !}.
    \end{aligned}
\end{equation}
As noted in the discussion of the tadpole contribution, similarly here, $k$ should be viewed as the sum of the incoming momenta at vertex $x$ and the outgoing ones at $y$, and therefore $k\ll M$.Expanding in $k^2/M^2$ then gives the leading contribution given above.\\
\begin{figure}[tbp]
\centering 
\subfigure[]{\includegraphics[width=.2\textwidth]{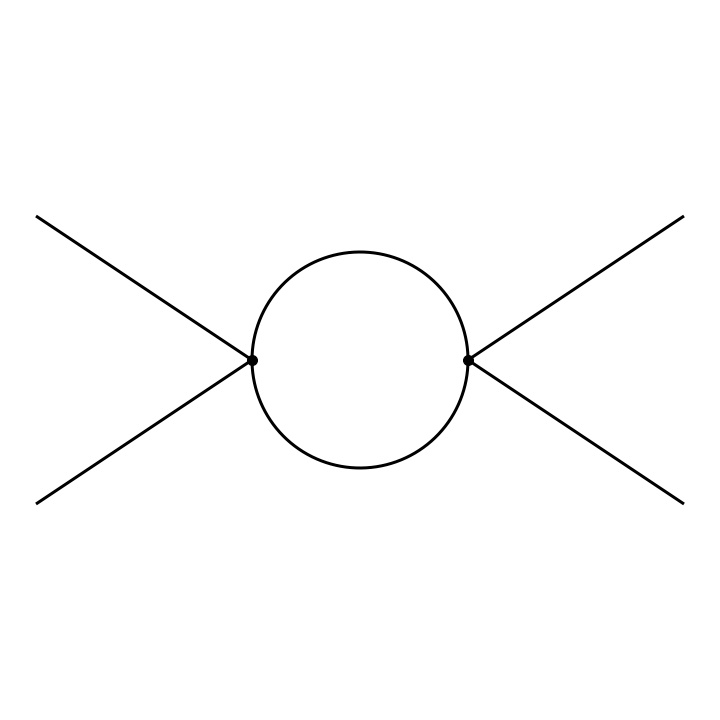}\label{fig:ch33a}}
\hfill
\subfigure[]{\includegraphics[width=.2\textwidth]{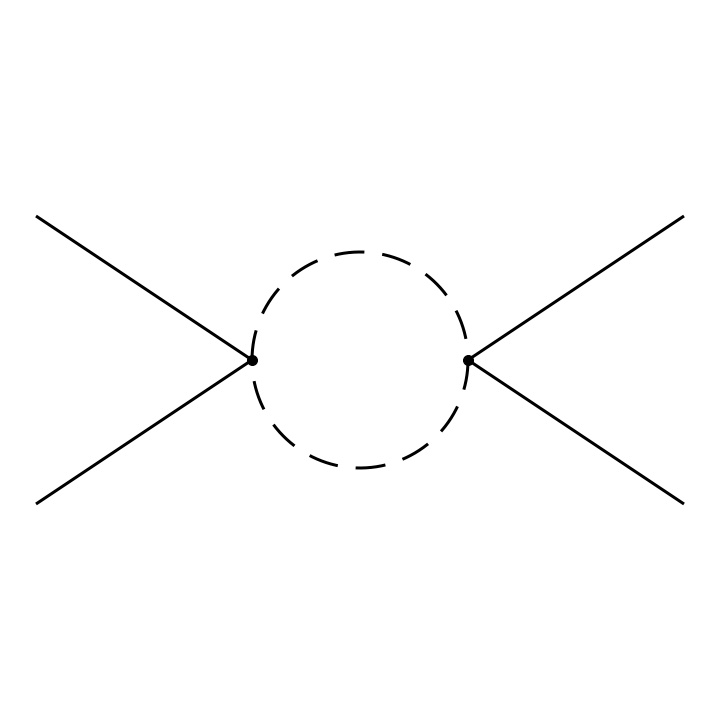}\label{fig:ch33b}}
\hfill
\subfigure[]{\includegraphics[width=.2\textwidth]{Figures/4c.jpg}\label{fig:ch33c}}
\hfill
\subfigure[]{\includegraphics[width=.2\textwidth]{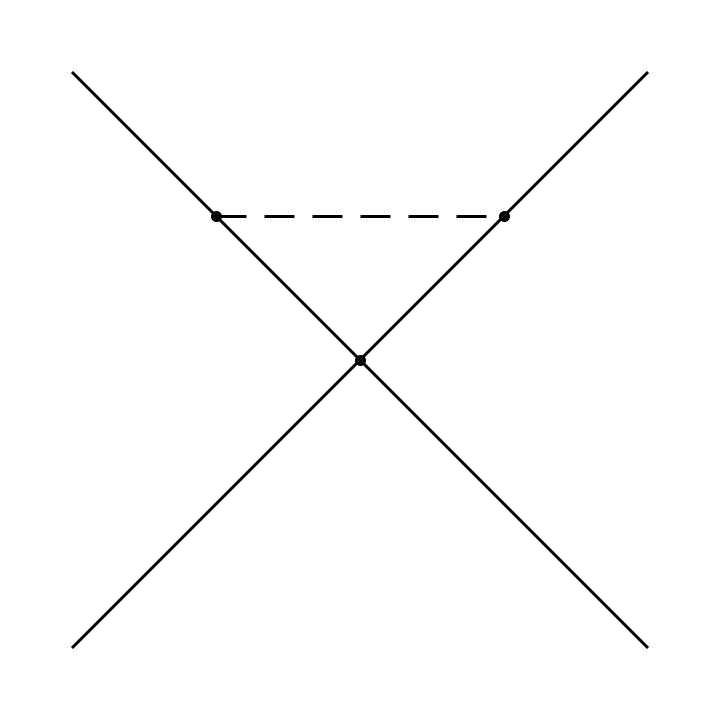}\label{fig:ch33d}}
\caption{\label{fig:i} Four point functions at one loop level.}
\end{figure}
\indent In loop calculation using dimensional regularization, we will again encounter extra finite terms proportional to $C=-\int_{k<M}\frac{d^{3}k}{(2\pi)^{3}}\,\frac{1}{\sqrt{k^{2}+m^{2}}}$, similar to those we encountered in section \ref{subsubsec:ch321}. However, in the interest of making our expressions appear more compact, we will omit writing them down explicitly.We will confirm in section \ref{sec:EFT} that all large log terms $\ln{\frac{m^{2}}{M^{2}}}$ produced in the four point function calculation are canceled out when constructing the effective Hamiltonian, which implies that these extra finite terms like $C$ will be canceled as well. Furthermore, to simplify calculations, we will set the total incoming and outgoing external momenta to zero.\\
\indent Let's consider Figure \ref{fig:ch33a}. This diagram is of order $\lambda_{0}^{2}$, and it arises from the term $\frac{i}{2} \left[H_{A}^{4,1},\Omega_{0}^{4,1}\right]$ where $H_{A}^{4,1}$ comes from the mode expansion of $\frac{\lambda_0}{4!}\Phi_{L}^{4}(x)$. In this expansion we must pick a term of form $\phi^{2}\chi^{2}$ which gives
\begin{equation}\label{HA}
    \begin{aligned} H_{A}^{4,1}= \int d^{3}x \,  \frac{\lambda_{0}}{4} \phi^{2}(x) \sum_{k}\sum_{p} \frac{1}{2\sqrt{\epsilon_{p}\epsilon_{k}}}\left(b_{p} b_{k} e^{i\left(\boldsymbol p + \boldsymbol k\right)\boldsymbol x}+b_{p}^{\dagger}b_{k}^{\dagger}e^{-i\left(\boldsymbol p + \boldsymbol k\right)\boldsymbol x}\right).
    \end{aligned}
\end{equation}
From the equation $i\left[H_{2},\Omega_{0}^{4,1}\right]+H_{A}^{4,1}=0$, we can get the corresponding $\Omega_{0}^{4,1}$ to be
\begin{equation}\label{omegaA}
    \begin{aligned} \Omega_{0}^{4,1} = \int d^{3}y \, \frac{\lambda_{0}}{4} \phi^{2}(y) \sum_{k^{'}} \sum_{p^{'}} \left(\frac{-i e^{i\left(\boldsymbol k^{'}+\boldsymbol p^{'}\right)\boldsymbol y}}{2\sqrt{\epsilon_{k^{'}}\epsilon_{p^{'}}}\left(\epsilon_{k^{'}}+\epsilon_{p^{'}}\right)}b_{k^{'}}b_{p^{'}}+\frac{i e^{-i\left(\boldsymbol k^{'}+\boldsymbol p^{'}\right)\boldsymbol y}}{2\sqrt{\epsilon_{k^{'}}\epsilon_{p^{'}}}\left(\epsilon_{k^{'}}+\epsilon_{p^{'}}\right)}b_{k^{'}}^{\dagger}b_{p^{'}}^{\dagger}\right).
    \end{aligned}
\end{equation}
Then $\frac{i}{2} \left[H_{A}^{4,1},\Omega_{0}^{4,1}\right]$ yields
\begin{equation}\label{eq:3.29}
    \begin{aligned}  -\frac{3\lambda_{0}^{2}}{32\pi^{2}}\left(\frac{1}{\overline{\epsilon}}-\ln\frac{m^{2}}{\mu^{2}}\right)\int d^{3}x \, \frac{\phi^{4}(x)}{4!}.
    \end{aligned}
\end{equation}
In Eq.(\ref{HA}) and Eq.(\ref{omegaA}) and subsequently their commutator, we may view $\boldsymbol k +\boldsymbol p$ as the total incoming momentum since that is the momentum entering the vertex at $x$. In evaluating the commutator we have set this to zero which then
implies $\epsilon_{k} = \epsilon_{p} = \sqrt{k^{2}+m^{2}}$.\\ 
\indent The next contribution is from Figure \ref{fig:ch33b}. The calculation is similar to the first one except the corresponding $H_{A}$ is different. In this case we have 

\begin{subequations}\label{eq:y}
\begin{align}
\label{eq:y:1}
&H_{A}^{4,2}= \int d^{3}x \,  \frac{\lambda_{2}}{4} \phi^{2}(x) \sum_{k}\sum_{p} \frac{1}{2\sqrt{\omega_{p}\omega_{k}}}\left(a_{p} a_{k} e^{i\left(\boldsymbol p + \boldsymbol k\right)\boldsymbol x}+a_{p}^{\dagger}a_{k}^{\dagger}e^{-i\left(\boldsymbol p + \boldsymbol k\right)\boldsymbol x}\right),\\
\label{eq:y:2}
&\Omega_{0}^{4,2} = \int d^{3}y \, \frac{\lambda_{2}}{4} \phi^{2}(y) \sum_{k^{'}} \sum_{p^{'}} \left(\frac{-i e^{i\left(\boldsymbol k^{'}+\boldsymbol p^{'}\right)\boldsymbol y}}{2\sqrt{\omega_{k^{'}}\omega_{p^{'}}}\left(\omega_{k^{'}}+\omega_{p^{'}}\right)}a_{k^{'}}a_{p^{'}}+\frac{i e^{-i\left(\boldsymbol k^{'}+\boldsymbol p^{'}\right)\boldsymbol y}}{2\sqrt{\omega_{k^{'}}\omega_{p^{'}}}\left(\omega_{k^{'}}+\omega_{p^{'}}\right)}a_{k^{'}}^{\dagger}a_{p^{'}}^{\dagger}\right).
\end{align}
\end{subequations}
The commutator is then evaluated as:
\begin{equation}\label{eq:x}
    \begin{aligned} \frac{i}{2} \left[H_{A}^{4,2},\Omega_{0}^{4,2}\right] = -\frac{3\lambda_{2}^{2}}{32\pi^{2}}\left(\frac{1}{\overline{\epsilon}}-\ln\frac{M^{2}}{\mu^{2}}\right)\int d^{3}x \, \frac{\phi^{4}(x)}{4!}.
    \end{aligned}
\end{equation}
\begin{table}[t]
    \centering
    \begin{tabular}{|lll|} 
    \hline
    $H_{A}^{4,3}$ &$\Omega_{0}^{4,3,1}$& $\Omega_{0}^{4,3,2}$ \\ \hline  
    First combination (a) & & \\ 
    $\lambda_{1}a\phi^{2}$ &  $\lambda_{1}a^{\dagger}b^{\dagger}b^{\dagger}$ & $\lambda_{0}bb\phi^{2}$\\
    $\lambda_{1}a^{\dagger}\phi^{2}$ & $\lambda_{1}abb$ & $\lambda_{0}b^{\dagger}b^{\dagger}\phi^{2}$\\ \hline
    Second combination (b) & & \\
    $\lambda_{1}abb$ & $ \lambda_{1}a^{\dagger}\phi^{2}$ & $\lambda_{0}b^{\dagger}b^{\dagger}\phi^{2}$\\
    $\lambda_{1}a^{\dagger}b^{\dagger}b^{\dagger}$ & $\lambda_{1}a\phi^{2}$ & $ \lambda_{0}bb\phi^{2}$\\ \hline
    Third combination (c) & &\\
    $\lambda_{1}abb$ & $\lambda_{0}b^{\dagger}b^{\dagger}\phi^{2}$ & $\lambda_{1}a^{\dagger}\phi^{2}$\\
    $\lambda_{1}a^{\dagger}b^{\dagger}b^{\dagger}$ & $\lambda_{0}bb\phi^{2}$ & $\lambda_{1}a\phi^{2}$\\ \hline
    Fourth combination (d) & &\\
    $\lambda_{0}bb\phi^{2}$ & $\lambda_{1}a^{\dagger}b^{\dagger}b^{\dagger}$ & $\lambda_{1}a\phi^{2}$\\
    $\lambda_{0}b^{\dagger}b^{\dagger}\phi^{2}$ & $\lambda_{1}abb$ & $\lambda_{1}a^{\dagger}\phi^{2}$ \\ \hline
    \end{tabular}
    \caption{Combinations of $-\frac{1}{3}\left[\left[H_{A}^{4,3},\Omega_{0}^{4,3,1}\right],\Omega_{0}^{4,3,2}\right]$ that contribute to Figure \ref{fig:ch33c}. Various entries are identified only by coupling constant and operator structure.}
    \label{tab:ch31}
\end{table}
\indent Next we calculate the contributions that correspond to Figure \ref{fig:ch33c}.
Notice that, in this diagram, we have the product of coupling constants as $\lambda_{0}\lambda_{1}^{2}$, which means we need both $-\frac{1}{3}\left[\left[H_{A},\Omega_{0}\right],\Omega_{0}\right]$ and $-\frac{1}{2}\left[\left[H_{B},\Omega_{0}\right],\Omega_{0}\right]$ from Eq.(\ref{Hdecoup}). For consistency of notation we label $H_{A}$ in the first commutator as $H_{A}^{4,3}$, the first $\Omega_{0}$ next to $H_{A}^{4,3}$ as $\Omega_{0}^{4,3,1}$, the second $\Omega_{0}$ as $\Omega_{0}^{4,3,2}$ and $H_{B}$ as $H_{B}^{4,3}$. There are many ways to pick $H_{A}^{4,3}$, $\Omega_{0}^{4,3,1}$, and $\Omega_{0}^{4,3,2}$ in $-\frac{1}{3} \left[\left[H_{A}^{4,3},\Omega_{0}^{4,3,1}\right],\Omega_{0}^{4,3,2}\right]$, and we split them into four kinds of combinations shown in Table \ref{tab:ch31}. We will explicitly show the calculation of the first combination in Table \ref{tab:ch31}, i.e., (a)  while a detailed discussion of contributions of the others is relegated to appendix \ref{sec:app32}.\\
\indent Consider the following terms from the first combination in Table \ref{tab:ch31}, i.e., (a):
\begin{subequations}\label{eq:y}
\begin{align}
\label{eq:y:1}
H_{A}^{4,3} &= \frac{\lambda_{1}}{2} \int d^{3}x \, \sum_{k} \frac{\phi^{2}(x)}{\sqrt{2\omega_{k}}} a_{k}e^{i \boldsymbol k \boldsymbol x},
\\
\label{eq:y:2}
\Omega_{0}^{4,3,1} &= \frac{\lambda_{1}}{2} \int d^{3}y \, \sum_{k^{'}} \sum_{p} \sum_{q} 
\frac{i e^{-i\left(\boldsymbol k^{'} + \boldsymbol p + \boldsymbol q\right) \boldsymbol y}}{2^{\frac{3}{2}}\sqrt{\omega_{k^{'}}\epsilon_{p}\epsilon_{q}}\left(\omega_{k^{'}}+\epsilon_{p}+\epsilon_{q}\right)}a_{k^{'}}^{\dagger}b_{p}^{\dagger}b_{q}^{\dagger},
\\
\label{eq:y:3}
\Omega_{0}^{4,3,2} &= \frac{\lambda_{0}}{4} \int d^{3}z \, \sum_{p^{'}} \sum_{q^{'}} \phi^{2}(z) \frac{-i e^{i\left(\boldsymbol p^{'} +\boldsymbol q^{'}\right)\boldsymbol z}}{2\sqrt{\epsilon_{p^{'}}\epsilon_{q^{'}}}\left(\epsilon_{p^{'}}+\epsilon_{q^{'}}\right)}b_{p^{'}}b_{q^{'}}.
\end{align}
\end{subequations}
We first calculate $\left[H_{A}^{4,3},\Omega_{0}^{4,3,1}\right]$
\begin{equation}\label{eq:x}
    \begin{aligned} \left[H_{A}^{4.3},\Omega_{0}^{4,3,1}\right] &= \frac{\lambda_{1}^{2}}{4} \int d^{3}x \int d^{3}y \, \phi^{2}(x) \sum_{k}\sum_{k^{'}}\sum_{p}\sum_{q} \frac{i e^{i \boldsymbol k \boldsymbol x} e^{-i\left(\boldsymbol k^{'} + \boldsymbol p + \boldsymbol q\right) \boldsymbol y} }{2^{2}\sqrt{\omega_{k}\omega_{k^{'}}\epsilon_{p}\epsilon_{q}}\left(\omega_{k^{'}}+\epsilon_{p}+\epsilon_{q}\right)}\\& \left[a_{k},a_{k^{'}}^{\dagger}b_{p}^{\dagger}b_{q}^{\dagger}\right],\\
    & = \frac{\lambda_{1}^{2}}{4} \int d^{3}x \int d^{3}y \, \phi^{2}(x)\sum_{k}\sum_{p}\sum_{q} \frac{i e^{i \boldsymbol k \left(\boldsymbol x - \boldsymbol y\right)} e^{-i\left( \boldsymbol p + \boldsymbol q\right) \boldsymbol y} }{2^{2}\omega_{k}\sqrt{\epsilon_{p}\epsilon_{q}}\left(\omega_{k^{}}+\epsilon_{p}+\epsilon_{q}\right)} b_{p}^{\dagger} b_{q}^{\dagger},\\
    & = \frac{i\lambda_{1}^{2}}{16} \int d^{3}x \, \phi^{2}(x) \sum_{p} \sum_{q} \frac{e^{-i\left(\boldsymbol p + \boldsymbol q\right)\boldsymbol x }}{M\sqrt{\epsilon_{p}\epsilon_{q}}\left(M+\epsilon_{p}+\epsilon_{q}\right)}b^{\dagger}_{p}b^{\dagger}_{q}.
    \end{aligned}
\end{equation}
In the above, we have put the total external momentum to zero, which implies the momentum $k$ associated with heavy particle is therefore 0 and $w_{k} = M$.
Next we have
\begin{equation}\label{eq:x}
    \begin{aligned} -\frac{1}{3}\left[\left[H_{A}^{4,3},\Omega_{0}^{4,3,1}\right],\Omega_{0}^{4,3,2}\right] &= -\frac{\lambda_{1}^{2}\lambda_{0}}{192} \int d^{3}x \int d^{3}z \,\sum_{p}\sum_{q}\sum_{p^{'}}\sum_{q^{'}} \phi^{2}(x)\phi^{2}(z)\\
    & \frac{e^{i\left(\boldsymbol p^{'} + \boldsymbol q^{'}\right)\boldsymbol z -i\left(\boldsymbol p + \boldsymbol q\right)\boldsymbol x}}{2^{2}M\sqrt{\epsilon_{p}\epsilon_{q}\epsilon_{p^{'}}\epsilon_{q^{'}}}\left(\epsilon_{p^{'}}+\epsilon_{q^{'}}\right)\left(M+\epsilon_{p}+\epsilon_{q}\right)} \left[b_{p}^{\dagger}b_{q}^{\dagger},b_{p^{'}}b_{q^{'}}\right],\\
    &= \frac{\lambda_{1}^{2}\lambda_{0}}{192} \int d^{3}x \int d^{3}z \,\sum_{p} \sum_{q} \phi^{2}(x) \phi^{2}(z) \frac{e^{i\left(\boldsymbol p + \boldsymbol q\right)\left(\boldsymbol x - \boldsymbol z\right)}}{2^{2}M\epsilon_{p}^{3}\left(M+2\epsilon{p}\right)},\\
    &= \frac{\lambda_{1}^{2}\lambda_{0}}{192} \int d^{3}x \, \phi^{4}(x) \int\frac{d^{3}p}{\left(2\pi\right)^{3}}\,\frac{1}{2^{2}M\epsilon_{p}^{3}\left(M+2\epsilon_{p}\right)},
    \end{aligned}
\end{equation}
where we have used the fact that $\boldsymbol p +\boldsymbol q = \boldsymbol k = 0$, and thus $\epsilon_{p}=\epsilon_{q}$. Including the Hermitian conjugate, we have the result for combination Table \ref{tab:ch31}(a) to be,
\begin{equation}\label{eq:x}
    \frac{\lambda_{1}^{2}\lambda_{0}}{96}\int d^{3}x \, \phi^{4}(x) \int\frac{d^{3}p}{\left(2\pi\right)^{3}}\,\frac{1}{2^{2}M\epsilon_{p}^{3}\left(M+2\epsilon_{p}\right)}.
\end{equation}
\indent The combinations Table \ref{tab:ch31}(b) and Table \ref{tab:ch31}(c) will yield the same result:
\begin{equation}
    \begin{aligned} \frac{\lambda_{1}^{2}\lambda_{0}}{96}\int d^{3}x \, \phi^{4}(x) \int \frac{d^{3}p}{(2\pi)^{3}}\, \frac{1}{2M^{2}\epsilon_{p}^{3}}.
    \end{aligned}
\end{equation}
\indent and Table \ref{tab:ch31}(d) gives,
\begin{equation}
    \begin{aligned} \frac{\lambda_{1}^{2}\lambda_{0}}{96}\int d^{3}x \, \phi^{4}(x) \int \frac{d^{3}p}{(2\pi)^{3}}\, \frac{1}{M^{2}\epsilon_{p}^{2}(M+2\epsilon_{p})}.
    \end{aligned}
\end{equation}
\indent Adding all four kinds of combinations together, we will get 
\begin{equation}\label{eq:x}
    \begin{aligned} &\frac{\lambda_{1}^{2}\lambda_{0}}{96}\int d^{3}x \, \phi^{4}(x) \int \frac{d^{3}p}{\left(2\pi\right)^{3}} \, \left(\frac{1}{M^{2}\epsilon_{p}^{2}\left(M+2\epsilon_{p}\right)}+\frac{1}{2\epsilon_{p}^{3}M\left(M+2\epsilon_{p}\right)}+\frac{1}{M^{2}\epsilon_{p}^{3}}\right),\\
    &=\frac{\lambda_{1}^{2}\lambda_{0}}{64} \int d^{3}x \phi^{4}(x) \int \frac{d^{3}p}{\left(2\pi\right)^{3}}\,\frac{1}{M^{2}\epsilon_{p}^{3}},\\
    &= \frac{3\lambda_{1}^{2}\lambda_{0}}{32\pi^{2}M^{2}} \int d^{3}x \, \frac{\phi^{4}(x)}{4!} \left(\frac{1}{\overline{\epsilon}}-\ln\frac{m^{2}}{\mu^{2}}\right).
    \end{aligned}
\end{equation}
\begin{table}[t]
    \centering
    \begin{tabular}{|lll|} 
    \hline
    $H_{B}^{4,3}$ &$\Omega_{0}^{4,3,1}$& $\Omega_{0}^{4,3,2}$ \\ \hline  

    $\lambda_{1}b^{\dagger}b^{\dagger}a$ & $\lambda_{0}bb\phi^{2}$ & $\lambda_{1}a^{\dagger}\phi^{2}$ \\

    $\lambda_{1}b^{\dagger}b^{\dagger}a$ & $\lambda_{1}a^{\dagger}\phi^{2}$ & $\lambda_{0}bb\phi^{2}$ \\ 

    $\lambda_{1}a^{\dagger}bb$ & $\lambda_{1}a\phi^{2}$ & $\lambda_{0}b^{\dagger}b^{\dagger}\phi^{2}$ \\ 

    $\lambda_{1}a^{\dagger}bb$ & $\lambda_{0}b^{\dagger}b^{\dagger}\phi^{2}$ & $\lambda_{1}a\phi^{2}$ \\
    \hline
    \end{tabular}
    \caption{Combinations of $-\frac{1}{2}\left[\left[H_{B}^{4,3},\Omega_{0}^{4,3,1}\right],\Omega_{0}^{4,3,2}\right]$ that contribute to Figure \ref{fig:ch33c}.}
    \label{tab:ch32}
\end{table}
All possible combinations of $-\frac{1}{2}\left[\left[H_{B}^{4,3},\Omega_{0}^{4,3,1}\right],\Omega_{0}^{4,3,2}\right]$ are shown in Table \ref{tab:ch32}, and the result is 
\begin{equation}
    \begin{aligned} \frac{3\lambda_{1}^{2}\lambda_{0}}{32\pi^{2}M^{2}} \int d^{3}x \, \frac{\phi^{4}(x)}{4!} \left(\frac{1}{\overline{\epsilon}}-\ln\frac{m^{2}}{\mu^{2}}\right).
    \end{aligned}
\end{equation}
Therefore, the total contribution from Figure \ref{fig:ch33c} is, 
\begin{equation}
    \begin{aligned} \frac{3\lambda_{1}^{2}\lambda_{0}}{16\pi^{2}M^{2}} \int d^{3}x \, \frac{\phi^{4}(x)}{4!} \left(\frac{1}{\overline{\epsilon}}-\ln\frac{m^{2}}{\mu^{2}}\right).
    \end{aligned}
\end{equation}
\begin{table}[t]
    \centering
    \begin{tabular}{|lll|} 
    \hline
    $H_{B}^{4,4}$ &$\Omega_{0}^{4,4,1}$& $\Omega_{0}^{4,4,2}$ \\ \hline  
    First combination (a) & & \\ 
    $\lambda_{0}b^{\dagger}b^{\dagger}\phi^{2}$ &  $\lambda_{1}a^{\dagger}b^{\dagger}\phi$ & $\lambda_{1}ab\phi$\\
    $\lambda_{0}b^{\dagger}b\phi^{2}$ & $\lambda_{1}ab\phi$ & $\lambda_{1}a^{\dagger}b^{\dagger}\phi$\\ \hline
    Second combination (b) & & \\
    $\lambda_{1}b^{\dagger}a\phi$ & $ \lambda_{1}a^{\dagger}b^{\dagger}\phi$ & $\lambda_{0}bb\phi^{2}$\\
    $\lambda_{1}a^{\dagger}b\phi$ & $\lambda_{1}ab\phi$ & $ \lambda_{0}b^{\dagger}b^{\dagger}\phi^{2}$\\ \hline
    Third combination (c) & &\\
    $\lambda_{1}b^{\dagger}a\phi$ & $\lambda_{0}bb\phi^{2}$ & $\lambda_{1}a^{\dagger}b^{\dagger}\phi$\\
    $\lambda_{1}a^{\dagger}b^{\dagger}\phi$ & $\lambda_{0}b^{\dagger}b^{\dagger}\phi^{2}$ & $\lambda_{1}ab\phi$\\ \hline
    \end{tabular}
    \caption{Combinations of $-\frac{1}{2}\left[\left[H_{B}^{4,4},\Omega_{0}^{4,4,1}\right],\Omega_{0}^{4,4,2}\right]$ that contribute to Figure \ref{fig:ch33c}.}
    \label{tab:ch33}
\end{table}
\indent As for Figure \ref{fig:ch33d}, only the term  $-\frac{1}{2}\left[\left[H_{B}^{4,4},\Omega_{0}^{4,4,1}\right],\Omega_{0}^{4,4,2}\right]$ will contribute. Similarly, we divide the whole commutator into several kinds of combinations shown in Table \ref{tab:ch33} and calculate each separately. The final results are given below and details may be found in appendix \ref{sec:app33}.\\
\indent Table \ref{tab:ch33}(a) will give
\begin{equation}\label{eq:x}
    \begin{aligned} \frac{\lambda_{0}\lambda_{1}^{2}}{16} \int d^{3}x  \, \phi^{4}(x) \int \frac{d^{3}k}{\left(2\pi\right)^{3}} \frac{1}{\omega_{k}\epsilon_{k}^{2}\left(\omega_{k}+\epsilon_{k}\right)^{2}}.
    \end{aligned}
\end{equation}
\indent Since Table \ref{tab:ch33}(b) and Table \ref{tab:ch33}(c) only differ from an exchange of $\Omega_{0}^{4,4,1}$ and $\Omega_{0}^{4,4,2}$, they will give the same result:
\begin{equation}\label{eq:x}
    \begin{aligned} \frac{\lambda_{1}^{2}\lambda_{0}}{32} \int d^{3}x \, \phi^{4}(x) \int \frac{d^{3}p}{\left(2\pi\right)^{3}} \, \frac{1}{\omega_{p}\epsilon_{p}^{3}\left(\omega_{p}+\epsilon_{p}\right)}.
    \end{aligned}
\end{equation}
\indent Adding the contributions from three combinations we get,
\begin{equation}\label{eqfig5d}
    \begin{aligned} -\frac{1}{2}\left[\left[H_{B}^{4,4},\Omega_{0}^{4,4,1}\right],\Omega_{0}^{4,4,2}\right] = \frac{\lambda_{1}^{2}\lambda_{0}}{16} \int d^{3}x \, \phi^{4}(x) \int \frac{d^{3}k}{\left(2\pi\right)^{3}} \, \left[\frac{1}{\omega_{k}\epsilon_{k}^{2}\left(\omega_{k}+\epsilon_{k}\right)^{2}}+\frac{1}{\omega_{k}\epsilon_{k}^{3}\left(\omega_{k}+\epsilon_{k}\right)}\right].
    \end{aligned}
\end{equation}
Let's consider $\int \frac{d^{3}k}{\left(2\pi\right)^{3}} \frac{1}{\omega_{k}\epsilon_{k}^{3}\left(\omega_{k}+\epsilon_{k}\right)} $ first,
\begin{equation}\label{eq:x}
    \begin{aligned} \int \frac{d^{3}k}{\left(2\pi\right)^{3}}\, \frac{1}{\omega_{k}\epsilon_{k}^{3}\left(\omega_{k}+\epsilon_{k}\right)} &=  \int \frac{d^{3}k}{\left(2\pi\right)^{3}} \, \left(\frac{1}{\epsilon^{3}_{k}\left(M^{2}-m^{2}\right)}-\frac{1}{\omega_{k}\epsilon_{k}^{2}\left(\omega_{k}+\epsilon_{k}\right)\left(\omega_{k}-\epsilon_{k}\right)}\right),\\
    &\approx \int \frac{d^{3}k}{\left(2\pi\right)^{3}} \, \left(\frac{1}{\epsilon^{3}_{k}M^{2}}-\frac{1}{\omega_{k}\epsilon_{k}^{2}\left(\omega_{k}+\epsilon_{k}\right)\left(\omega_{k}-\epsilon_{k}\right)}\right).
    \end{aligned}
\end{equation}
Combining the last term with the first one in in Eq.(\ref{eqfig5d}), we get,
\begin{equation}\label{eq:x}
    \begin{aligned} &\int \frac{d^{3}k}{\left(2\pi\right)^{3}} \, \left(\frac{1}{\omega_{k}\epsilon_{k}^{2}\left(\omega_{k}+\epsilon_{k}\right)^{2}}-\frac{1}{\omega_{k}\epsilon_{k}^{2}\left(\omega_{k}+\epsilon_{k}\right)\left(\omega_{k}-\epsilon_{k}\right)}\right),\\
    &= \int \frac{d^{3}k}{\left(2\pi\right)^{3}} \, \frac{-2\epsilon_{k}}{\omega_{k}\epsilon_{k}^{2}\left(\omega_{k}+\epsilon_{k}\right)^{2}\left(\omega_{k}-\epsilon_{k}\right)},\\
    &=\int \frac{d^{3}k}{\left(2\pi\right)^{3}} \, \frac{-2\epsilon_{k}\left(\omega_{k}-\epsilon_{k}\right)}{\omega_{k}\epsilon_{k}^{2}\left(\omega_{k}+\epsilon_{k}\right)^{2}\left(\omega_{k}-\epsilon_{k}\right)^{2}},\\
    &\approx\int \frac{d^{3}k}{\left(2\pi\right)^{3}} \, \left(\frac{-2}{\epsilon_{k}M^{4}}+\frac{2}{\omega_{k}M^{4}}\right).
    \end{aligned}
\end{equation}
To the order $O(\frac{1}{M^{2}})$, we only need to keep $\int \frac{d^{3}k}{\left(2\pi\right)^{3}} \frac{-2}{\omega_{k}M^{4}} $. The net contribution from Figure \ref{fig:ch33d} is then,
\begin{equation}\label{eq:x}
    \begin{aligned} &\frac{\lambda_{1}^{2}\lambda_{0}}{16} \int d^{3}x\, \phi^{4}(x) \int \frac{d^{3}k}{\left(2\pi\right)^{3}} \left(\frac{2}{\omega_{k}M^{4}}+\frac{1}{\epsilon_{k}^{3}M^{2}}\right),\\
    & =-\frac{\lambda_{1}^{2}\lambda_{0}}{64\pi^{2}M^{2}}\int d^{3}x\, \phi^{4}(x) \left(\left(\frac{1}{\overline{\epsilon}}+1-\ln\frac{M^{2}}{\mu^{2}}\right)-\left(\frac{1}{\overline{\epsilon}}-\ln\frac{m^{2}}{\mu^{2}}\right)\right)m,\\
    &= -\frac{3\lambda_{0}\lambda_{1}^{2}}{8\pi^{2}M^{2}}\left(1-\ln\frac{M^{2}}{\mu^{2}}+\ln\frac{m^{2}}{\mu^{2}}\right)\int d^{3}x \, \frac{\phi^{4}(x)}{4!}.
    \end{aligned}
\end{equation}
\indent Taking $\mu \approx M$ and summing the contributions from the four diagrams together, we get the final result of four point functions at one loop level, up to order $O(\frac{1}{M^{2}})$:
\begin{equation}
    \begin{aligned} &-\frac{3\lambda_{0}^{2}}{32\pi^{2}}\left(\frac{1}{\overline{\epsilon}}-\ln\frac{m^{2}}{M^{2}}\right)\int d^{3}x \, \frac{\phi^{4}(x)}{4!}-\frac{3\lambda_{2}^{2}}{32\pi^{2}}\frac{1}{\overline{\epsilon}}\int d^{3}x \, \frac{\phi^{4}(x)}{4!}\\
    &+\frac{3\lambda_{1}^{2}\lambda_{0}}{16\pi^{2}M^{2}}\left(\frac{1}{\overline{\epsilon}}-\ln\frac{m^{2}}{M^{2}}\right)\int d^{3}x \, \frac{\phi^{4}(x)}{4!}-\frac{3\lambda_{1}^{2}\lambda_{0}}{8\pi^{2}M^{2}}\left(1+\ln\frac{m^{2}}{M^{2}}\right)\int d^{3}x \, \frac{\phi^{4}(x)}{4!}.
    \end{aligned}
\end{equation}
\subsection{Renormalization}
\label{subsec:ch33}
We argued in section \ref{subsec:ch23} that the UV divergence in the calculation of $H_{decoupled}$ contains information regarding the renormalization of $H_{1}$, where $H_{1}$ is just the low energy part in the full Hamiltonian. Let's first write $H_{1}$ as
\begin{equation}
    \begin{aligned} H_{1}=\int d^{3}x\,\frac{1}{2}(\Pi_{bare}^{2}(x)+(\nabla \phi_{bare}(x))^{2})+\frac{1}{2}m_{bare}^{2}\phi_{bare}^{2}(x)+\frac{\lambda_{0}^{bare}}{4 !}\phi_{bare}^{4}(x),
    \end{aligned}
\end{equation}
and then introduce the renormalization Z factor such that
\begin{subequations}\label{eq:y}
\begin{align}
\label{eq:y:1}
& \phi_{bare}=\sqrt{Z_{\phi}}\phi,\\
\label{eq:y:2}
& \Pi_{bare}=\frac{\Pi}{\sqrt{Z_{\phi}}},\\
\label{eq:y:3}
&m_{bare}=\sqrt{Z_{m}}m,\\
\label{eq:y:4}
&\lambda_{0}^{bare}=Z_{\lambda_{0}}\lambda_{0}.
\end{align}
\end{subequations}
Expanding these Z factors in terms of $\frac{1}{\overline{\epsilon}}$, we have
\begin{subequations}\label{eq:y}
\begin{align}
\label{eq:y:1}
&Z_{\phi}=1+\delta^{1}_{\phi}(\frac{1}{\overline{\epsilon}})+O(\frac{1}{\overline{\epsilon}^{2}}),\\
\label{eq:y:2}
&Z_{m}=1+\delta^{1}_{m}(\frac{1}{\overline{\epsilon}})+O(\frac{1}{\overline{\epsilon}^{2}}),\\
\label{eq:y:3}
&Z_{\lambda_{0}}=1+\delta^{1}_{\lambda_{0}}(\frac{1}{\overline{\epsilon}})+O(\frac{1}{\overline{\epsilon}^{2}}).
\end{align}
\end{subequations}
Implementing these expansion, we can rewrite $H_{1}$ as
\begin{equation}
    \begin{aligned}
    H_{1}=&\int d^{3}x\, \frac{1}{2}(\Pi^{2}(x)+(\nabla\phi(x))^{2})+\frac{1}{2}m^{2}\phi^{2}(x)+\frac{\lambda_{0}}{4 !}\phi^{2}(x)\\
    & \frac{1}{2}\delta_{\phi}^{1}(-\Pi^{2}(x)+(\nabla\phi(x))^{2})+\frac{1}{2}(\delta_{\phi}^{1}+\delta_{m}^{1})m^{2}\phi^{2}(x)+\frac{\lambda_{0}}{4 !}(2\delta_{\phi}^{1}+\delta_{\lambda_{0}}^{1})\phi^{4}(x)+O(\frac{1}{\overline{\epsilon}^{2}}).
    \end{aligned}
\end{equation}
Using $\overline{MS}$ scheme, we can cancel the UV divergence in $H_{decoupled}$ by counterterms in $H_{1}$. From two point calculation in section \ref{subsubsec:ch321}, we have divergent terms:
\begin{equation}
    \begin{aligned} \left(-\frac{\lambda_{0}m^{2}}{32\pi^{2}}-\frac{\lambda_{2}M^{2}}{32\pi^{2}}+\frac{\lambda_{1}^{2}m^{2}}{32\pi^{2}M^{2}}-\frac{\lambda_{1}^{2}}{16\pi^{2}}\right)\frac{1}{\overline{\epsilon}}\int d^{3}x\, \frac{\phi^{2}(x)}{2}.
    \end{aligned}
\end{equation}
From four point calculation in section \ref{subsubsec:ch322}, we have divergent terms:
\begin{equation}
    \begin{aligned}
    \left(-\frac{3\lambda_{0}^{2}}{32\pi^{2}}-\frac{3\lambda_{2}^{2}}{32\pi^{2}}+\frac{3\lambda_{1}^{2}\lambda_{0}}{16\pi^{2}M^{2}}\right)\frac{1}{\overline{\epsilon}}\int d^{3}x\,\frac{\phi^{4}(x)}{4 !}.
    \end{aligned}
\end{equation}
Since, there is no term proportional to the momentum, we know $\delta_{\phi}^{1} = 0$. Then we can use the following renormalization conditions:
\begin{subequations}
\begin{align}
&\left(\delta_{m}^{1}m^{2}+\left(-\frac{\lambda_{0}m^{2}}{32\pi^{2}}-\frac{\lambda_{2}M^{2}}{32\pi^{2}}+\frac{\lambda_{1}^{2}m^{2}}{32\pi^{2}M^{2}}-\frac{\lambda_{1}^{2}}{16\pi^{2}}\right)\frac{1}{\overline{\epsilon}}\right)\int d^{3}x\, \frac{\phi^{2}(x)}{2}=0,\\
&\left(\delta_{\lambda_{0}}^{1}\lambda_{0}+\left(-\frac{3\lambda_{0}^{2}}{32\pi^{2}}-\frac{3\lambda_{2}^{2}}{32\pi^{2}}+\frac{3\lambda_{1}^{2}\lambda_{0}}{16\pi^{2}M^{2}}\right)\frac{1}{\overline{\epsilon}}\right)\int d^{3}x\, \frac{\phi^{4}(x)}{4 !}=0,
\end{align}
\end{subequations}
to obtain
\begin{subequations}
\begin{align}
    &\delta_{\phi}^{1}=0,\\
    &\delta_{m}^{1}=\frac{1}{m^{2}}\left(\frac{\lambda_{0}m^{2}}{32\pi^{2}}+\frac{\lambda_{2}M^{2}}{32\pi^{2}}-\frac{\lambda_{1}^{2}m^{2}}{32\pi^{2}M^{2}}+\frac{\lambda_{1}^{2}}{16\pi^{2}}\right)\frac{1}{\overline{\epsilon}},\\
    &\delta_{\lambda_{0}}^{1}=\frac{1}{\lambda_{0}}\left(\frac{3\lambda_{0}^{2}}{32\pi^{2}}+\frac{3\lambda_{2}^{2}}{32\pi^{2}}-\frac{3\lambda_{1}^{2}\lambda_{0}}{16\pi^{2}M^{2}}\right)\frac{1}{\overline{\epsilon}},
\end{align}
\end{subequations}
which agree with the results from the traditional implementation of renormalization in the Lagrangian framework.
After renormalization, we get the decoupled Hamiltonian at one loop order to be
\begin{equation}\label{eq:3.56}
    \begin{aligned} H_{decoupled}^{one loop}= &\int d^{3}x\, \frac{1}{2}(\Pi^{2}(x)+(\nabla\phi(x))^{2})+\frac{1}{2}m^{2}\phi^{2}(x)+\frac{\lambda_{0}}{4 !}\phi^{2}(x)\\
    & (\frac{\lambda_{0}m^{2}}{32\pi^{2}}(\ln{\frac{m^{2}}{M^{2}}}-1)-\frac{\lambda_{2}M^{2}}{32\pi^{2}}+\frac{3\lambda_{1}^{2}m^{2}}{32\pi^{2}M^{2}}(-\ln{\frac{m^{2}}{M^{2}}}+1)\\
    &-\frac{\lambda_{1}^{2}}{16\pi^{2}}(1+\ln{\frac{m^{2}}{M^{2}}})+(\frac{\lambda_{0}}{4}-\frac{3\lambda_{1}^{2}}{2M^{2}})C)\frac{\phi^{2}(x)}{2}+\frac{\lambda_{1}^{2}}{32\pi^{2}M^{2}}\frac{1}{2}(\Pi^{2}(x)+(\nabla\phi(x))^{2})\\
    &+(\frac{3\lambda_{0}^{2}}{32\pi^{2}}\ln{\frac{m^{2}}{M^{2}}}-\frac{3\lambda_{1}^{2}\lambda_{0}}{16\pi^{2}M^{2}}\ln{\frac{m^{2}}{M^{2}}}-\frac{3\lambda_{1}^{2}\lambda_{0}}{8\pi^{2}M^{2}}(1+\ln{\frac{m^{2}}{M^{2}}}))\frac{\phi^{4}(x)}{4!},
    \end{aligned}
\end{equation}
where we again omit the term proportional to $C=-\int_{k<M} \frac{d^{3} k}{(2 \pi)^{3}} \frac{1}{\sqrt{k^{2}+m^{2}}}$ in the four point function calculation.

\section{Construction of the Effective Field Theory}
\label{sec:EFT}
In this section we will use the results from section \ref{sec:Calc} to construct the one loop matched effective field theory up to $O(\frac{1}{M^{2}})$. We will also check that the matching corrections are analytic in low energy parameters. This is essential for the overall consistency of the program and requires the cancellation during matching of the infrared contributions in $H_{decoupled}$ like the $\ln{\frac{m^{2}}{M^{2}}}$ terms and those labelled by $C$. \\
\begin{figure}[t]
\centering 
\subfigure[]{\includegraphics[width=.25\textwidth]{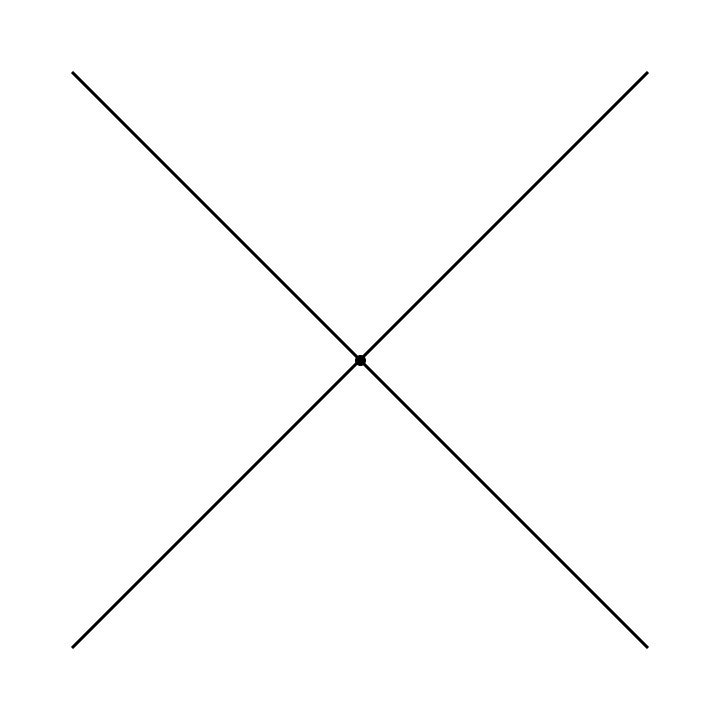}\label{fig:ch41a}}
\hfill
\subfigure[]{\includegraphics[width=.25\textwidth]{Figures/4pt2.jpg}\label{fig:ch41b}}
\hfill
\caption{\label{fig:i} Tree level contributions in $H_{decoupled}$.}
\end{figure}
\indent At tree level up to order $O(\frac{1}{M^{2}})$, we have
\begin{equation}
            H_{decoupled}^{tree}=\int d^{3}x\, \frac{1}{2}(\Pi^{2}(x)+(\nabla\phi(x))^{2}+\frac{1}{2}m^{2}\phi^{2}(x)+(\lambda_{0}-\frac{3\lambda_{1}}{M^{2}})\frac{\phi^{4}(x)}{4 !}+O(\frac{1}{M^{4}}),
\end{equation}
that corresponds to Figure \ref{fig:ch41a} and Figure \ref{fig:ch41b}. First, projecting the decoupled Hamiltonian at tree level onto the low energy subspace, we have
\begin{equation}\label{eq:4.2}
    \begin{aligned} H_{eff}^{tree} = H_{decoupled}^{tree} = \int d^{3}x\, \frac{1}{2}(\Pi^{2}(x)+(\nabla\phi(x))^{2})+\frac{1}{2}m^{2}\phi^{2}(x)+(\lambda_{0}-\frac{3\lambda_{1}}{M^{2}})\frac{\phi^{4}(x)}{4 !}+O(\frac{1}{M^{4}}).
    \end{aligned}
\end{equation}
In order to obtain the physical effective Hamiltonian defined in the complete light field Hilbert space and to include loop corrections, we need to switch the low frequency modes $\phi$'s in Eq.(\ref{eq:4.2}) back to the full light field $\Phi_{L}$. Then we can decouple the high frequency and low frequency modes by unitary transformations:
\begin{equation}
    \begin{aligned} H_{decoupled}^{'}=\bra{0_{H}}\omega^{'\dagger}H_{eff}^{tree}\omega^{'}\ket{0_{H}},
    \end{aligned}
\end{equation}
where $H_{decoupled}^{'}$ is the decoupled Hamiltonian corresponding to $H_{eff}^{tree}$ and $\omega^{'}$ is the unitary transformation. Similar to what we did in section \ref{subsec:ch21}, we decompose $\omega^{'}$ into a series of unitary transformations
\begin{equation}
    \omega^{'}=\omega_{0}^{'}\omega_{1}^{'}\omega_{2}^{'}...,
\end{equation}
and each $\omega_{i}$ can be further written as $\omega'_{i}=e^{i\Omega'_{i}}$. The function of this series of unitary transformations $\omega^{'}$ is to disentangle the high energy states above scale $M$ from the low energy ones order by order. Furthermore, we decompose the effective Hamiltonian at tree level into four parts, analogous to the decomposition in section \ref{subsec:ch21},
\begin{equation}
H_{eff}^{tree}=H_{eff\,1}^{tree}+H_{eff\,2}^{tree}+H_{eff\,A}^{tree}+H_{eff\,B}^{tree},    
\end{equation}
where $H_{eff\,1}^{tree}$ only contains the low frequency modes, $H_{eff\,2}^{tree}$ is the free part for high frequency modes, $H_{eff\,A}^{tree}$ contains terms that only have high frequency annihilation or creation operators, and $H_{eff\,B}^{tree}$ denotes whatever is left over.
Starting from the zeroth order transformation $\omega_{0}^{'}$, we have 
\begin{equation}
    \label{eq:x}
    \begin{aligned} &\omega_{0}^{'\dagger}\left(H_{eff\,1}^{tree}+H_{eff\,2}^{tree}+H_{eff\,A}^{tree}+H_{eff\,B}^{tree}\right)\omega_{0}^{'}\\
    =&e^{-i\Omega_{0}^{'}}\left(H_{eff\,1}^{tree}+H_{eff\,2}^{tree}+H_{eff\,A}^{tree}+H_{eff\,B}^{tree}\right)e^{i\Omega_{0}^{'}}\\
    =&H_{eff\,1}^{tree}+H_{eff\,2}^{tree}+H_{eff\,A}^{tree}+H_{eff\,B}^{tree}\\
    &+i[H_{eff\,1}^{tree},\Omega_{0}^{'}]+i[H_{eff\,2}^{tree},\Omega_{0}^{'}]+i[H_{eff\,A}^{tree},\Omega_{0}^{'}]+i[H_{eff\,B}^{tree},\Omega_{0}^{'}]...
    \end{aligned}
\end{equation}    
We can then eliminate $H_{eff\,A}^{tree}$ by imposing the decoupling condition at zeroth order:
\begin{equation}
    i\left[H_{eff\,2}^{tree},\Omega^{'}_{0}\right]+H_{eff\,A}^{tree}=0.
\end{equation}
Next, we go to first order and eliminate the new term $i\left[H_{eff\,1}^{tree},\Omega^{'}_{0}\right]$ generated in the zeroth order decoupling that contains only annihilation or creation operators by imposing
\begin{equation}
    i\left[H_{eff\,1}^{tree},\Omega^{'}_{0}\right]+i\left[H_{eff\,2}^{tree},\Omega^{'}_{1}\right]=0.
\end{equation}
Our result for the decoupled effective Hamiltonian at first order then is,
\begin{equation}
    \begin{aligned} H_{decoupled}^{'}=&\bra{0_{H}} H_{eff\,1}^{tree}+H_{eff\,2}^{tree}+H_{eff\,B}^{tree}+\frac{i}{2}[H_{eff\,A}^{tree},\Omega^{'}_{0}]+i[H_{eff\,B}^{tree},\Omega_{0}^{'}]+... \ket{0_{H}},
    \end{aligned}
\end{equation}
which is good enough for constructing the effective field theory up to $O(\frac{1}{M^{2}})$.\\
\begin{figure}[t]
\centering 
\includegraphics[width=.35\textwidth]{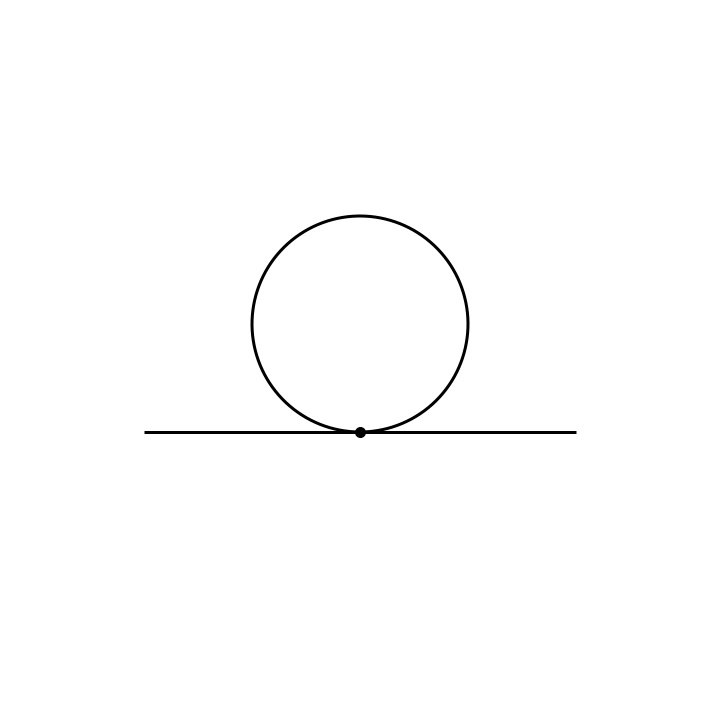}
\caption{\label{fig:ch42} One loop two point function contribution in $H_{decoupled}^{'}$}
\end{figure}
\indent Let us begin with the two point function shown in Figure \ref{fig:ch42}, which arises from the normal ordering of term $(\lambda_{0}-\frac{3\lambda_{1}}{M^{2}})\frac{\Phi_{L}^{4}(x)}{4 !}$ in $H_{eff}^{tree}$. The calculation here is similar to the calculation of Figure \ref{fig:ch31a}, and we only need to substitute $\lambda_{0}$ with $\lambda_{0}-\frac{3\lambda_{1}^{2}}{M^{2}}$. Therefore, the result is 
\begin{equation}
    \begin{aligned} -\frac{m^{2}}{32\pi^{2}}(\lambda_{0}-\frac{3\lambda_{1}^{2}}{M^{2}})(\frac{1}{\overline{\epsilon}}-\ln{\frac{m^{2}}{\mu^{2}}}+1)\int d^{3}x \, \frac{\phi^{2}(x)}{2}+\frac{\lambda_{0}-\frac{3\lambda_{1}^{2}}{M^{2}}}{4}C\int d^{3}x\, \frac{\phi^{2}(x)}{2},
    \end{aligned}
\end{equation}
where $C=-\int_{k<M}\frac{d^{3}k}{(2\pi)^{3}}\,\frac{1}{\sqrt{k^{2}+m^{2}}}$ is the same as in section \ref{subsubsec:ch321}. Then we do the renormalization in $\overline{MS}$ scheme to cancel out the divergent part and take $\mu \approx M$. The finite terms left are,
\begin{equation}
    \begin{aligned} -\frac{m^{2}}{32\pi^{2}}(\lambda_{0}-\frac{3\lambda_{1}^{2}}{M^{2}})(-\ln{\frac{m^{2}}{M^{2}}}+1)\int d^{3}x \, \frac{\phi^{2}(x)}{2}+\frac{\lambda_{0}-\frac{3\lambda_{1}^{2}}{M^{2}}}{4}C\int d^{3}x\, \frac{\phi^{2}(x)}{2}.
    \end{aligned}
\end{equation}
\begin{figure}[t]
\centering 
\includegraphics[width=.35\textwidth]{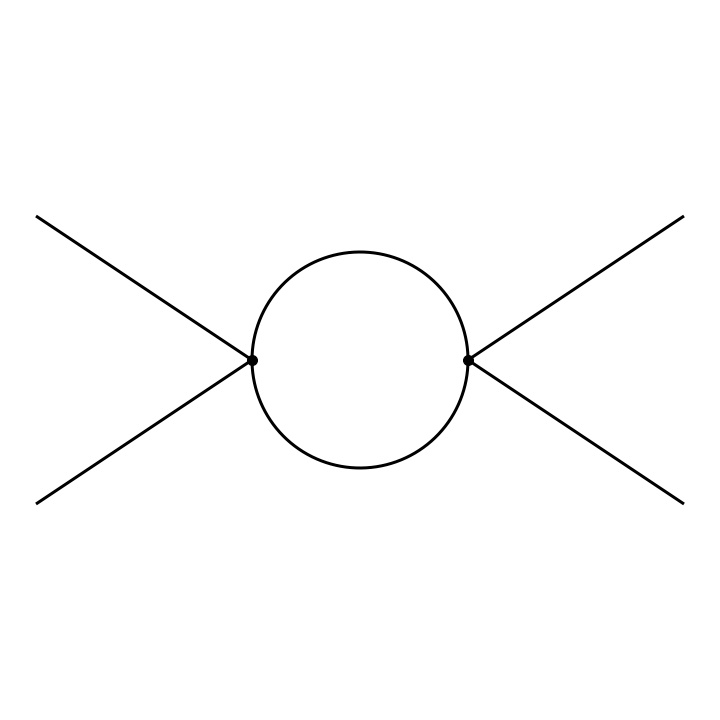}
\caption{\label{fig:ch43} One loop four point function contribution in $H_{decoupled}^{'}$}
\end{figure}
\indent We next consider the contribution from the four point function shown in Figure \ref{fig:ch43}. This arises from the commutator $\frac{i}{2}[H_{eff\,A}^{tree},\Omega^{'}_{0}]$, where,
\begin{subequations}\label{eq:y}
\begin{align}
\label{eq:y:1}
H_{eff\,A}^{tree}&= \int d^{3}x \,  \frac{\lambda_{0}-\frac{3\lambda_{1}^{2}}{M^{2}}}{4} \phi^{2}(x) \sum_{k}\sum_{p} \frac{1}{2\sqrt{\epsilon_{p}\epsilon_{k}}}\left(b_{p} b_{k} e^{i\left(\boldsymbol p + \boldsymbol k\right)\boldsymbol x}+b_{p}^{\dagger}b_{k}^{\dagger}e^{-i\left(\boldsymbol p + \boldsymbol k\right)\boldsymbol x}\right),
\\
\label{eq:y:2}
\Omega_{0}^{'} &= \int d^{3}y \,\frac{\lambda_{0}-\frac{3\lambda_{1}^{2}}{M^{2}}}{4} \phi^{2}(y) \sum_{k^{'}} \sum_{p^{'}} \left(\frac{-i e^{i\left(\boldsymbol k^{'}+\boldsymbol p^{'}\right)\boldsymbol y}}{2\sqrt{\epsilon_{k^{'}}\epsilon_{p^{'}}}\left(\epsilon_{k^{'}}+\epsilon_{p^{'}}\right)}b_{k^{'}}b_{p^{'}}+\frac{i e^{-i\left(\boldsymbol k^{'}+\boldsymbol p^{'}\right)\boldsymbol y}}{2\sqrt{\epsilon_{k^{'}}\epsilon_{p^{'}}}\left(\epsilon_{k^{'}}+\epsilon_{p^{'}}\right)}b_{k^{'}}^{\dagger}b_{p^{'}}^{\dagger}\right).
\end{align}
\end{subequations}
Again the calculation is similar to the one of Figure \ref{fig:ch33a}, and we only need to substitute $\lambda_{0}$ with $\lambda_{0}-\frac{3\lambda_{1}^{2}}{M^{2}}$. Hence, we have
\begin{equation}
    \begin{aligned} -\frac{3\left(\lambda_{0}-\frac{3\lambda_{1}^{2}}{M^{2}}\right)^{2}}{32\pi^{2}}\left(\frac{1}{\overline{\epsilon}}-\ln{\frac{m^{2}}{M^{2}}}\right)\int d^{3}x\, \frac{\phi^{4}(x)}{4 !},
    \end{aligned}
\end{equation}
where we have omitted the finite terms proportional to C in the four point calculation as we did in section \ref{subsubsec:ch322}.\\
After renormalization, up to $O(\frac{1}{M^{2}})$, we are left with
\begin{equation}
    \begin{aligned} \left(\frac{3\lambda_{0}^{2}}{32\pi^{2}}-\frac{9\lambda_{1}^{2}\lambda_{0}}{16\pi^{2}M^{2}}\right)\ln{\frac{m^{2}}{M^{2}}}\int d^{3}x\,\frac{\phi^{4}(x)}{4 !}.
    \end{aligned}
\end{equation}

Putting the results from two and four point calculation together, we get
\begin{equation}\label{eq:4.15}
    \begin{aligned} H_{decoupled}^{oneloop,eff}=&\int d^{3}x\, \frac{1}{2}(\Pi^{2}(x)+(\nabla\phi(x))^{2})+\frac{1}{2}m^{2}\phi^{2}(x)\\&+(-\frac{m^{2}}{32\pi^{2}}(\lambda_{0}-\frac{3\lambda_{1}^{2}}{M^{2}})(-\ln{\frac{m^{2}}{M^{2}}}+1)+\frac{\lambda_{0}-\frac{3\lambda_{1}^{2}}{M^{2}}}{4}C)\frac{\phi^{2}(x)}{2}\\&+(\lambda_{0}-\frac{3\lambda_{1}}{M^{2}}+(\frac{3\lambda_{0}}{32\pi^{2}}-\frac{9\lambda_{1}^{2}\lambda_{0}}{16\pi^{2}M^{2}})\ln{\frac{m^{2}}{M^{2}}})\frac{\phi^{4}(x)}{4 !}.
    \end{aligned}
\end{equation}
\indent To get the effective theory at one loop order, we simply subtract Eq.(\ref{eq:4.15}) from Eq.(\ref{eq:3.56}) and then switch $\phi$ back to $\Phi_{L}$. Thus, to $O(\frac{1}{M^{2}})$, the effective Hamiltonian at one loop so constructed in this theory is:
\begin{equation}
    \begin{aligned} H_{eff}^{one loop}=&\int d^{3}x\, \frac{1}{2}(\Pi_{L}^{2}+(\nabla\Phi_{L}(x))^{2})+\frac{1}{2}m^{2}\Phi_{L}^{2}(x)-\left(\frac{\lambda_{2}M^{2}}{32\pi^{2}}+\frac{\lambda_{1}^{2}}{16\pi^{2}}(1+\frac{m^{2}}{M^{2}})\right)\frac{\Phi_{L}^{2}(x)}{2}\\
    &+\left(\lambda_{0}-\frac{3\lambda_{1}^{2}}{M^{2}}-\frac{3\lambda_{0}\lambda_{1}^{2}}{8\pi^{2}M^{2}}\right)\frac{\Phi_{L}^{4}(x)}{4 !}+\frac{\lambda_{1}^{2}}{32\pi^{2}M^{2}} \frac{1}{2}(\Pi^{2}_{L}(x)+\nabla\Phi_{L}(x))^{2}).
    \end{aligned}
\end{equation}
As promised, the extra finite terms proportional to $C=-\int_{k<M} \frac{d^{3} k}{(2 \pi)^{3}} \frac{1}{\sqrt{k^{2}+m^{2}}}$ have all canceled out along with the large log terms proportional to $\ln{\frac{m^{2}}{M^{2}}}$ during matching. The final result agrees with \cite{Rothstein:2003mp} where starting from the same full theory an effective lagrangian was obtained using standard methods.\\

\section{The Renormalization Group Equations}
\label{sec:Renormgroup}
In this section we first obtain the exact renormalization group equations in the context of the renormalization scheme advocated here and apply it at the perturbative level to the scalar field theory example. For simplicity of presentation we will limit the application to the pure $\lambda \phi^4$ theory.As in the previous sections, we will be working in the Hamiltonian framework at fixed time.

As we have emphasized in this paper and explicitly shown in the scalar field theoretical example of the previous sections, the process of renormalization and decoupling of heavy particle effects can be regarded as the result of unitary transformations which decouple the entanglement between the low and high momentum modes of an interacting field theory. The transformed Hamiltonian then incorporates effects of the high energy modes on the low energy physics.In this section, we will generically denote by $\mu$ the cut-off scale separating the low momentum modes from those at high momenta.
Let the disentanglement of these low and high momentum modes in the Hilbert space be implemented by the unitary transformation $\omega(\mu)$ whose action on the states is given by:
\begin{equation}
    \ket{\Psi(\mu)} = \omega^{\dagger}(\mu)\ket{\Psi},
\end{equation}
and under a change in the cut-off scale, 
\begin{equation}
    {\partial \over \partial \mu}\ket{\Psi(\mu)} = G_{\mu}\ket{\Psi(\mu)},
\end{equation}
with the generator of scale transformations identified as $G_{\mu}=({\partial\omega^{\dagger} \over \partial \mu})\omega$.As discussed earlier, the corresponding change in the Hamiltonian when the unitary transformations are time-independent is 
\begin{equation}\label{Hamil-change}
    H'(\mu)=\omega(\mu)^{\dagger}H\omega(\mu),
\end{equation}
where $H$ denotes the scale independent starting Hamiltonian. Taking the derivative of both sides of Eq.(\ref{Hamil-change}) with respect to $\mu$, we obtain,
\begin{equation}\label{RG-1}
    {\partial H' \over \partial\mu} = [G_{\mu} , H'].
\end{equation}
The renormalization group equation follows from the observation that the change in the Hamiltonian with scale is compensated by the corresponding changes in the coupling parameters of the theory, i.e.,
\begin{equation}\label{RG-2}
    {\partial m^2 \over \partial \mu}{\partial H' \over \partial m^2}+
    {\partial \lambda \over \partial \mu}{\partial H' \over \partial \lambda} = - [G_{\mu} , H']
\end{equation}
This is an exact equation which is not very useful in practical calculations. The procedure, which is more appropriate for a perturbative expansion, that we have followed in the previous sections is to project the Hamiltonian $H'$ on to the high energy vacuum and to expand the right hand side of Eq.(\ref{RG-2}). We will implement this below at the one-loop level using the results from the previous sections as needed. By expanding, $\omega=e^{i\Omega_0}$, we get,
\begin{equation}
G_{\mu} = -i{\partial \Omega_0 \over \partial \mu}+\frac{1}{2}[{\partial \Omega_0 \over \partial \mu}, \Omega_0] + ....
\end{equation}
This can now be used to evaluate the expansion of the right hand side of Eq.(\ref{RG-2}). For $H'$ we will use the expansion Eq.(\ref{Hdecoup}), and keeping only the terms which are needed here we get,
\begin{equation}
  [G_{\mu} , H'] = \frac{i}{2}{\partial \over \partial \mu}[H_A, \Omega_0]
  + ...
\end{equation}
In obtaining these we have used the fact that to this order $H_A$ is independent of $\mu$.
Let us apply this next to the four point function and obtain the $\beta$ function for the coupling constant $\lambda$. 

Since we are working to the one loop order, the contribution on the left hand side of Eq.(\ref{RG-2}) comes from the term ${\lambda \over 4!}\int d^3x \phi^4$ in $H'$ and on the right hand side we have the $\mu$ dependent contribution from Eq.(\ref{eq:3.29})(after $\overline{MS}$ renormalization),
\begin{equation}
    \frac{i}{2}[H_A^{4,1}, \Omega_0^{4,1}]= {-3\lambda^2 \over 32\pi^2}\ln{{\mu^2 \over m^2}}\int d^3x {\phi^4 \over 4!}.
\end{equation}
Putting this together, we finally get the well known result,
\begin{equation}
    \mu{\partial \lambda \over \partial \mu} = \beta(\lambda) = {3\lambda^2 \over 16\pi^2}.
\end{equation}
This provides a consistency check of our approach. The above procedure can, in principle, be extended iteratively to higher orders, however,at the expense of growing tedium. In order to get some new information, it would be interesting to extend this analysis to explore the decoupling at different momentum scales. Work in this direction is in progress.
\section{Discussion}
\label{sec:discussion}
In this paper we have shown the consistency of a Hamiltonian  renormalization framework which emphasizes its basic origin as due to the momentum space entanglement between the various modes of a quantum field theory. Using unitary transformations on states to decouple the high energy modes from the low energy ones and projecting the transformed Hamiltonian to the low energy subspace, correctly accounts for renormalization effects and the property of decoupling in quantum field theories. We have also shown how the same approach can be consistently used in the construction of effective field theories. Novel renormalization group equations were also obtained and shown to lead to beta functions which are consistent with more conventional approaches. The next step would be to understand how different measures of entanglement like entanglement entropy and mutual information (for a review see \cite{Headrick:2018ctr}) may be used to analyze the properties of decoupling and to shed light on another striking property of quantum field theories, namely the insensitivity of the low energy physics to the details of the short distance structure. Taking inspiration from the Ryu-Takayanagi formula \cite{Ryu:2006ef}, another related future direction could be to look for a possible role of geometry in the renormalization program.

\acknowledgments
Bingzheng Han gratefully acknowledges a summer fellowship awarded by the Leinweber Center for Theoretical Physics.

\appendix
\section{Appendix for Section 3} 
\label{sec:app3}
\subsection{Explicit Calculation of Equation \ref{eq:3.23}}\label{sec:app31}
We start from this integral:\\
\begin{equation}\label{eq:x}
    \begin{aligned} \frac{i}{2}\left[H_{A}^{2,2},\Omega_{0}^{2,2}\right]&= - \int \int d^{3}x d^{3}y \, \sum_{k} \sum_{M<p} \lambda_{1}^{2} \frac{ \phi(x) \phi(y)}{4\omega_{k}\epsilon_{p}\left(\omega_{k}+\epsilon_{p}\right)} e^{i \left(\boldsymbol k + \boldsymbol p\right) \left(\boldsymbol x - \boldsymbol y\right)},\\
    &= - \int \int d^{3}x d^{3}y \, \sum_{k} \sum_{M<p} \lambda_{1}^{2} \frac{ \phi(x) \phi(y)e^{i \left(\boldsymbol k + \boldsymbol p\right) \left(\boldsymbol x - \boldsymbol y\right)}}{4\sqrt{k^{2}+M^{2}}\sqrt{p^{2}+m^{2}}\left(\sqrt{k^{2}+M^{2}}+\sqrt{p^{2}+m^{2}}\right)}.
    \end{aligned}
\end{equation}
In order to calculate this complicated integral, we need to split the fraction into two parts:
\begin{equation}\label{eq:A2}
    \begin{aligned} & \frac{1}{\sqrt{k^{2}+M^{2}}\sqrt{p^{2}+m^{2}}\left(\sqrt{k^{2}+M^{2}}+\sqrt{p^{2}+m^{2}}\right)},\\
     =& \frac{\sqrt{k^{2}+M^{2}}-\sqrt{p^{2}+m^{2}}}{\sqrt{k^{2}+M^{2}}\sqrt{p^{2}+m^{2}}\left(k^{2}+M^{2}-p^{2}-m^{2}\right)},\\
     =& \frac{1}{\sqrt{p^{2}+m^{2}}\left(k^{2}+M^{2}-p^{2}-m^{2}\right)} - \frac{1}{\sqrt{k^{2}+M^{2}}\left(k^{2}+M^{2}-p^{2}-m^{2}\right)}.
    \end{aligned}
\end{equation}
Let $\boldsymbol k + \boldsymbol p = \boldsymbol r$. We know $\boldsymbol r$ is total external momentum, thus r is much smaller than M. Also since $\boldsymbol k - \boldsymbol p$ is of order $O(M)$, we conclude that $\left(k^{2}-p^{2}\right) \ll M^{2}$. Hence we can write:\\
\begin{equation}\label{eq:x}
    \begin{aligned} & \frac{1}{\sqrt{p^{2}+m^{2}}\left(k^{2}+M^{2}-p^{2}-m^{2}\right)},\\
     = &\frac{1}{\sqrt{p^{2}+m^{2}}M^{2}\left(1-\frac{m^{2}}{M^{2}}+\frac{k^{2}-p^{2}}{M^{2}}\right)},\\
     \simeq & \frac{1}{\sqrt{p^{2}+m^{2}}M^{2}}\left(1+\frac{m^{2}}{M^{2}}-\frac{k^{2}-p^{2}}{M^{2}}+\left(\frac{k^{2}-p^{2}}{M^{2}}\right)^{2}\right).
    \end{aligned}
\end{equation}
To the order $\sim O(\frac{1}{M^{2}})$ this gives\\
\begin{equation}\label{eq:x}
    \begin{aligned} &\int d^{3}x \frac{m^{2}\lambda_{1}^{2}}{16\pi^{2}M^{2}}\left(\frac{1}{\overline{\epsilon}}-\ln{\frac{m^{2}}{\mu^{2}}}+1\right) \frac{\phi^{2}(x)}{2}-\frac{\lambda_{1}^{2}C}{2M^{2}}\int d^{3}x\,\frac{\phi^{2}(x)}{2}.
    \end{aligned}
\end{equation}
\indent Similarly, the second term from Eq.(\ref{eq:A2}) gives
\begin{equation}\label{eq:A5}
    \begin{aligned} \frac{1}{\sqrt{k^{2}+M^{2}}M^{2}}\left(1+\frac{m^{2}}{M^{2}}-\frac{k^{2}-p^{2}}{M^{2}}+\left(\frac{k^{2}-p^{2}}{M^{2}}\right)^{2}\right).
    \end{aligned}
\end{equation}
Straightforwardly, the first two terms in Eq.(\ref{eq:A5}) give\\
\begin{equation}\label{eq:x}
    \begin{aligned} & \int \int d^{3}x d^{3} y \, \lambda_{1}^{2} \phi(x) \phi(y) \int \frac{d^{3}k}{\left(2\pi\right)^{3}} \int \frac{d^{3}r}{\left(2\pi\right)^{3}} \frac{1}{4 \sqrt{k^{2}+M^{2}}M^{2}} e^{i \boldsymbol r \left(\boldsymbol x - \boldsymbol y\right)} \left(1+\frac{m^{2}}{M^{2}}\right),\\
    & = -\int d^{3}x \frac{\lambda_{1}^{2}}{16\pi^{2}}\left(\frac{1}{\overline{\epsilon}}-\ln{\frac{M^{2}}{\mu^{2}}}+1\right) \left(1+\frac{m^{2}}{M^{2}}\right) \frac{\phi^{2}(x)}{2}.
    \end{aligned}
\end{equation}
The last two terms are a bit tricky to handle. First, we note that $ k^{2}-p^{2} = \left(\boldsymbol k + \boldsymbol p\right) \left(\boldsymbol k -\boldsymbol p\right)= 2\boldsymbol k \boldsymbol r - r^{2}$. Then we get from the $\frac{k^{2}-p^{2}}{M^{2}}$ term:\\
\begin{equation}\label{eq:x}
    \begin{aligned}& \int \int d^{3}x d^{3} y \, \lambda_{1}^{2} \phi(x) \phi(y) \int \frac{d^{3}k}{\left(2\pi\right)^{3}} \int \frac{d^{3}r}{\left(2\pi\right)^{3}} \frac{1}{4 \sqrt{k^{2}+M^{2}}M^{2}} e^{i \boldsymbol r \left(\boldsymbol x - \boldsymbol y\right)} \left(-\frac{k^{2}-p^{2}}{M^{2}}\right),\\
    & = \int \int d^{3}x d^{3} y \, \lambda_{1}^{2} \phi(x) \phi(y) \int \frac{d^{3}k}{\left(2\pi\right)^{3}} \int \frac{d^{3}r}{\left(2\pi\right)^{3}} \frac{1}{4 \sqrt{k^{2}+M^{2}}M^{2}} e^{i \boldsymbol r \left(\boldsymbol x - \boldsymbol y\right)} \left(\frac{-2 \boldsymbol k \boldsymbol r +r^{2}}{M^{2}}\right),\\
    & =  \int \int d^{3}x d^{3} y \, \lambda_{1}^{2} \phi(x) \phi(y) \int \frac{d^{3}k}{\left(2\pi\right)^{3}} \int \frac{d^{3}r}{\left(2\pi\right)^{3}} \frac{1}{4 \sqrt{k^{2}+M^{2}}M^{2}} e^{i \boldsymbol r \left(\boldsymbol x - \boldsymbol y\right)}\frac{r^{2}}{M^{2}},\\
    & = \int \int d^{3}x d^{3}y \, \frac{\lambda_{1}^{2}}{4} \phi(x) \phi(y) \int \frac{d^{3}k}{\left(2\pi\right)^{3}} \frac{1}{M^{2}\sqrt{k^{2}+M^{2}}} \int \frac{d^{3}r}{\left(2\pi\right)^{3}} r^{2} e^{i \boldsymbol r \left(\boldsymbol x - \boldsymbol y\right)},\\
    & = -\int \int d^{3}x d^{3}y \, \frac{\lambda_{1}^{2}}{4} \phi(x) \phi(y) \int \frac{d^{3}k}{\left(2\pi\right)^{3}} \frac{1}{M^{2}\sqrt{k^{2}+M^{2}}} \left( \nabla_{y}^{2}\delta^{3}(\boldsymbol x -\boldsymbol y)\right),\\
    & = -\int d^{3}x \, \frac{\lambda_{1}^{2}}{4} \phi(x) \nabla^{2} \phi(x) \int \frac{d^{3}k}{\left(2\pi\right)^{3}} \frac{1}{M^{2}\sqrt{k^{2}+M^{2}}},\\
    & = -\int d^{3}x \, \frac{\lambda_{1}^{2}}{16\pi^{2}M^{2}}\left(\frac{1}{\overline{\epsilon}}-\ln{\frac{M^{2}}{\mu^{2}}}+1\right) \frac{1}{2}(\nabla \phi(x))^{2}.
    \end{aligned}
\end{equation}
To the $O(\frac{1}{M^{2}})$, we also need to consider the $\frac{4\left(\boldsymbol k \boldsymbol r\right)^{2}}{M^{4}}$ term from $\left(\frac{k^{2}-p^{2}}{M^{2}}\right)^{2}$:
\begin{equation}\label{eq:x}
    \begin{aligned} & \int \int d^{3}x d^{3} y \, \lambda_{1}^{2} \phi(x) \phi(y) \int \frac{d^{3}k}{\left(2\pi\right)^{3}} \int \frac{d^{3}r}{\left(2\pi\right)^{3}} \frac{1}{4 \sqrt{k^{2}+M^{2}}M^{2}} e^{i \boldsymbol r \left(\boldsymbol x - \boldsymbol y\right)}\frac{4\left(\boldsymbol k \boldsymbol r\right)^{2}}{M^{4}},\\ 
    & = \int \int d^{3}x d^{3} y \, \lambda_{1}^{2} \phi(x) \phi(y) \int \frac{d^{3}k}{\left(2\pi\right)^{3}} \frac{1}{ \sqrt{k^{2}+M^{2}}M^{6}} \int \frac{d^{3}r}{\left(2\pi\right)^{3}} \left(\boldsymbol k \boldsymbol r\right)^{2} e^{i \boldsymbol r \left(\boldsymbol x - \boldsymbol y\right)},\\
    & =  \int \int d^{3}x d^{3} y \, \lambda_{1}^{2} \phi(x) \phi(y) \int \frac{d^{3}r}{\left(2\pi\right)^{3}} r^{2} e^{i \boldsymbol r \left(\boldsymbol x - \boldsymbol y\right)} \int \frac{d^{3}k}{\left(2\pi\right)^{3}} \frac{k^{2}\cos^{2}{\theta}}{\sqrt{k^{2}+M^{2}}M^{6}},\\
    & = \int \int d^{3}x d^{3} y \, \lambda_{1}^{2} \phi(x) \phi(y) \int \frac{d^{3}r}{\left(2\pi\right)^{3}} r^{2} e^{i \boldsymbol r \left(\boldsymbol x - \boldsymbol y\right)} \frac{1}{32\pi^{2}M^{2}} \left(\frac{1}{\overline{\epsilon}}-\ln{\frac{M^{2}}{\mu^{2}}}+\frac{3}{2}\right),\\
    & = -\int \int d^{3}x d^{3} y \, \lambda_{1}^{2} \phi(x) \delta^{3}(\boldsymbol x- \boldsymbol y) \nabla_{y}^{2} \phi(y) \frac{1}{32\pi^{2}M^{2}} \left(\frac{1}{\overline{\epsilon}}-\ln{\frac{M^{2}}{\mu^{2}}}+\frac{3}{2}\right),\\
    & = \int d^{3}x\, \frac{\lambda_{1}^{2}}{16\pi^{2}M^{2}} \left(\frac{1}{\overline{\epsilon}}-\ln{\frac{M^{2}}{\mu^{2}}}+\frac{3}{2}\right) \frac{1}{2} (\nabla \phi(x))^{2}.
    \end{aligned}
\end{equation}
Putting together all the previous results, we can get\\
\begin{equation}\label{eq:x}
    \begin{aligned} 
    &\frac{m^{2}\lambda_{1}^{2}}{16\pi^{2}M^{2}}\left(\frac{1}{\overline{\epsilon}}-\ln{\frac{m^{2}}{\mu^{2}}}+1\right)\int d^{3}x\, \frac{\phi^{2}(x)}{2}-\frac{\lambda_{1}^{2}C}{2M^{2}}\int d^{3}x\,\frac{\phi^{2}(x)}{2}\\
    &-\frac{\lambda_{1}^{2}}{16\pi^{2}}\left(\frac{1}{\overline{\epsilon}}-\ln{\frac{M^{2}}{\mu^{2}}}+1\right)\left(1+\frac{m^{2}}{M^{2}}\right)\int d^{3}x \, \frac{\phi^{2}(x)}{2} +\frac{\lambda_{1}^{2}}{32\pi^{2}M^{2}}\int d^{3}x\,\frac{1}{2} (\nabla \phi(x))^{2}.
    \end{aligned}
\end{equation}
\subsection{Explicit Calculation of the Momentum Term in Equation \ref{eq:3.25}} \label{sec:app34}
As we mentioned in section \ref{subsubsec:ch321} the momentum term arises from $\bra{0_{H}}-\frac{1}{2}[[H_{1},\Omega_{0}],\Omega_{1}]\ket{0_{H}}$, we need to first find the corresponding $\Omega_{1}$. Since the momentum term is proportional to $\lambda_{1}^{2}\Pi^{2}$, we should determine $\Omega_{1}$ from 
\begin{equation}
    \begin{aligned} i[H_{1},\Omega_{0}^{2,2}]+i[H_{2},\Omega_{1}]=0.
    \end{aligned}
\end{equation}
It's straightforward to calculate $[H_{1},\Omega_{0}^{2,2}]$ using $[\phi(x),\Pi(y)]=i\delta^{3}(\boldsymbol x-\boldsymbol y)$, and the result is 
\begin{equation}\label{eq:A11}
    \begin{aligned} 
    \left[H_{1},\Omega_{0}^{2,2}\right] &= 2\int d^{3}x \int d^{3}y \int \sum_{k}\sum_{M<p} \left(\frac{\lambda_{1}}{2}\frac{-ia_{k}b_{p}e^{i(\boldsymbol p+\boldsymbol k)\boldsymbol y}}{2\sqrt{\omega_{k}\epsilon_{p}}(\omega_{k}+\epsilon_{p})}[\frac{1}{2}\Pi^{2}(x),\phi(y)]+h.c.\right),\\
    &=-\lambda_{1}\int d^{3}x \int \frac{d^{3}k}{(2\pi)^{3}} \int \frac{d^{3}p}{(2\pi)^{3}} \left(\frac{a_{k}b_{p}e^{i(\boldsymbol p+\boldsymbol k)\boldsymbol x}}{2\sqrt{\omega_{k}\epsilon_{p}}(\omega_{k}+\epsilon_{p})}\Pi(x)-h.c.\right).
    \end{aligned}
\end{equation}
Then $\Omega_{1}$ is calculated to be
\begin{equation}\label{eq:A12}
    \begin{aligned} 
    \Omega_{1}= -\int d^{3}y \int d^{3}q \int d^{3}r \left(\frac{\lambda_{1}a_{q}b_{r}e^{i(\boldsymbol q +\boldsymbol r)\boldsymbol y}}{2\sqrt{\omega_{q}\epsilon_{r}}(\omega_{q}+\epsilon_{r})^{2}}\Pi(y)+h.c.\right).
    \end{aligned}
\end{equation}
Using Eq.(\ref{eq:A11}) and Eq.(\ref{eq:A12}) we are able to calculate the commutator $-\frac{1}{2}[[H_{1},\Omega_{0}],\Omega_{1}]$. We will focus on commutator $[a_{k}b_{p},a^{\dagger}_{q}b^{\dagger}_{r}]$ since the other one will yield the same result.
\begin{equation}
    \begin{aligned} -\frac{1}{2}[[H_{1},\Omega_{0}],\Omega_{1}]&= -\frac{\lambda_{1}^{2}}{2}\int d^{3}x \int d^{3} y \int \frac{d^{3}k}{(2\pi)^{3}} \int \frac{d^{3}p}{(2\pi)^{3}} \int \frac{d^{3}q}{(2\pi)^{3}} \int \frac{d^{3}r}{(2\pi)^{3}} \\&\frac{\Pi(x)\Pi(y)e^{i(\boldsymbol k + \boldsymbol p)\boldsymbol x-i(\boldsymbol q +\boldsymbol r)\boldsymbol y}}{4 \sqrt{\omega_{k}\epsilon_{p}\omega_{q}\epsilon_{r}}(\omega_{k}+\epsilon_{p})(\omega_{q}+\epsilon_{r})^{2}}[a_{k}b_{p},a^{\dagger}_{q}b^{\dagger}_{r}],\\
    &= -\frac{\lambda_{1}^{2}}{8} \int d^{3}x \int d^{3}y \int \frac{d^{3}k}{(2\pi)^{3}} \int \frac{d^{3}p}{(2\pi)^{3}} \,\frac{e^{i(\boldsymbol p +\boldsymbol k)(\boldsymbol x -\boldsymbol y)}}{\omega_{k}\epsilon_{p}(\omega_{k}+\epsilon_{p})^{3}}\Pi(x)\Pi(y).
    \end{aligned}
\end{equation}
Here we have done normal ordering to get the final result.\\ 
\indent Similar to section \ref{sec:app31}, we can decompose the integral into four parts by multiplying both denominator and numerator by $(\omega_{k}-\epsilon_{p})^{3}$ to get 
\begin{equation}
    \begin{aligned} \frac{(\omega_{k}-\epsilon_{p})^{3}}{\omega_{k}\epsilon_{p}(k^{2}+M^{2}-p^{2}-m^{2})}=&\frac{\omega_{k}^{2}}{\epsilon_{p}M^{6}(1-\frac{m^{2}}{M^{2}}+\frac{k^{2}-p^{2}}{M^{2}})^{3}}-\frac{\epsilon_{p}^{2}}{\omega_{k}M^{6}(1-\frac{m^{2}}{M^{2}}+\frac{k^{2}-p^{2}}{M^{2}})^{3}}\\
    &  +\frac{3\epsilon_{p}}{M^{6}(1-\frac{m^{2}}{M^{2}}+\frac{k^{2}-p^{2}}{M^{2}})^{3}}-\frac{3\omega_{k}}{M^{6}(1-\frac{m^{2}}{M^{2}}+\frac{k^{2}-p^{2}}{M^{2}})^{3}}.
    \end{aligned}
\end{equation}
Since we are only interested in terms proportional to momentum $\Pi$, and it's from section \ref{sec:app31} that $\frac{k^{2}-p^{2}}{M^{2}}$ will mainly contribute to terms proportional to gradient, it's reasonable to make the assumption that $k\approx p$. Moreover up to order $O(M^{2})$, we can neglect the $m^{2}$ in $\epsilon_{p}$ and only need to consider the following three terms:
\begin{subequations}\label{eq:y}
\begin{align}
\label{eq:y:1}
&\frac{3\lambda_{1}^{2}}{8} \int d^{3}x \, \Pi^{2}(x) \int \frac{d^{3}k}{(2\pi)^{3}} \frac{\sqrt{k^{2}+M^{2}}}{M^{6}},
\\
\label{eq:y:2}
&\frac{\lambda_{1}^{2}}{8} \int d^{3}x \, \Pi^{2}(x) \int \frac{d^{3}k}{(2\pi)^{3}} \frac{k^{2}}{\sqrt{k^{2}+M^{2}}M^{6}},
\\
\label{eq:4:3}
&-\frac{\lambda_{1}^{2}}{8} \int d^{3}x \, \Pi^{2}(x) \int \frac{d^{3}k}{(2\pi)^{3}} \frac{k^{2}+M^{2}}{kM^{6}}.
\end{align}
\end{subequations}
The first term is simply calculated using dimensional regularization and the result is
\begin{equation}
    \begin{aligned} -\frac{3\lambda_{1}^{2}}{128M^{2}\pi^{2}} \left(\frac{1}{\overline{\epsilon}}-\ln\frac{M^{2}}{\mu^{2}}+\frac{3}{2}\right) \int d^{3}x \, \frac{1}{2}\Pi^{2}(x).
    \end{aligned}
\end{equation}
The second term can be calculated by splitting the integral into two terms:
\begin{equation}
    \begin{aligned} \frac{\lambda_{1}^{2}}{8} \int d^{3}x \, \Pi^{2}(x) \int \frac{d^{3}k}{(2\pi)^{3}} \frac{\sqrt{k^{2}+M^{2}}}{M^{6}}-\frac{\lambda_{1}^{2}}{8} \int d^{3}x \, \Pi^{2}(x) \int \frac{d^{3}k}{(2\pi)^{3}} \frac{1}{\sqrt{k^{2}+M^{2}}M^{4}}.
    \end{aligned}
\end{equation}
Again, using dimensional regularization we get:
\begin{equation}
    \begin{aligned} -\frac{\lambda_{1}^{2}}{128M^{2}\pi^{2}} \left(\frac{1}{\overline{\epsilon}}-\ln\frac{M^{2}}{\mu^{2}}+\frac{3}{2}\right) \int d^{3}x \, \frac{1}{2}\Pi^{2}(x)+\frac{\lambda_{1}^{2}}{32M^{2}\pi^{2}} \left(\frac{1}{\overline{\epsilon}}-\ln\frac{M^{2}}{\mu^{2}}+1\right) \int d^{3}x \, \frac{1}{2}\Pi^{2}(x).
    \end{aligned}
\end{equation}
The third term can be calculated as follows,
\begin{equation}
    \begin{aligned} &-\frac{\lambda_{1}^{2}}{8} \int d^{3}x \, \Pi^{2}(x) \int \frac{d^{3}k}{(2\pi)^{3}} \frac{k^{2}+M^{2}}{kM^{6}}\\
    =&-\frac{\lambda_{1}^{2}}{8M^{6}}\frac{2\pi^{\frac{3}{2}}}{\Gamma(\frac{3}{2})}\int \frac{dk}{(2\pi)^{3}}\,\frac{k}{(k^{2}+M^{2})^{-1}} \int d^{3}x \, \Pi^{2}(x),\\
    =&-\frac{\lambda_{1}^{2}}{8M^{6}}\frac{2\pi^{\frac{3}{2}}}{\Gamma(\frac{3}{2})}\frac{1}{\Gamma(-1+\delta)}\int \frac{dk}{(2\pi)^{3}}\,k \int d\lambda \, \lambda^{-2+\delta}e^{-\lambda(k^{2}+M^{2})}\int d^{3}x \, \Pi^{2}(x),\\
    =&-\frac{\lambda_{1}^{2}}{8M^{6}}\frac{2\pi^{\frac{3}{2}}}{\Gamma(\frac{3}{2})\Gamma(-1+\delta)}\frac{1}{(2\pi)^{3}}\int d\lambda \, \lambda^{-2+\delta} e^{-\lambda M^{2}} \int dk \,ke^{-\lambda k^{2}}\int d^{3}x \, \Pi^{2}(x),\\
    =&-\frac{\lambda_{1}^{2}}{16M^{6}}\frac{\pi^{\frac{3}{2}}}{\Gamma(\frac{3}{2})\Gamma(-1+\delta)}\frac{1}{(2\pi)^{3}}\int d\lambda \, \lambda^{-2+\delta} e^{-\lambda M^{2}} \int dk^{2} \,e^{-\lambda k^{2}}\int d^{3}x \, \Pi^{2}(x),\\
    =&-\frac{\lambda_{1}^{2}}{16M^{6}}\frac{\pi^{\frac{3}{2}}}{\Gamma(\frac{3}{2})\Gamma(-1+\delta)}\frac{1}{(2\pi)^{3}}\int d\lambda \, \lambda^{-3+\delta} e^{-\lambda M^{2}}\int d^{3}x \, \Pi^{2}(x),\\
    =&-\frac{\lambda_{1}^{2}}{16M^{6}}\frac{\pi^{\frac{3}{2}}\Gamma(-2+\delta)}{\Gamma(\frac{3}{2})\Gamma(-1+\delta)}\frac{M^{4}}{(2\pi)^{3}}\int d^{3}x \, \Pi^{2}(x),\\
    =&\frac{\lambda_{1}^{2}}{32M^{2}\pi^{2}}\int d^{3}x\,\frac{1}{2}\Pi^{2}(x).
    \end{aligned}
\end{equation}
In the above, $\delta$ is a regulator which is taken to zero at the end of the calculation, and we have used the standard representation:
\begin{equation*}
    {1 \over x^a} = \frac{1}{\Gamma(a)}\int d\lambda \lambda^{a-1}e^{-\lambda x}.
\end{equation*}
Putting results from all three terms together we get the contribution from the commutator $[a_{k}b_{p},a^{\dagger}_{q}b^{\dagger}_{r}]$ to be:
\begin{equation}
    \frac{\lambda_{1}^{2}}{64M^{2}\pi^{2}}\int d^{3}x\,\frac{1}{2}\Pi^{2}(x).
\end{equation}
Multiplying this result by 2 to take into account the contribution from commutator $[a_k^{\dagger}b_p^{\dagger},a_q b_r]$, we arrive at the final result for the canonical momentum piece in the decoupled Hamiltonian: 

\begin{equation}
    \frac{\lambda_{1}^{2}}{32M^{2}\pi^{2}}\int d^{3}x\,\frac{1}{2}\Pi^{2}(x).
\end{equation}
\subsection{Explicit Calculation of Figure \ref{fig:ch33c}}\label{sec:app32}
Since combination Table \ref{tab:ch31}(b) and Table \ref{tab:ch31}(c) only differ from an exchange of $\Omega_{0}^{4,3,1}$ and $\Omega_{0}^{4,3,2}$, they will yield the same result. Let's consider the following expansions for the quantities in Table \ref{tab:ch31}(b):
\begin{subequations}\label{eq:y}
\begin{align}
\label{eq:y:1}
& H_{A}^{4,3} = \frac{\lambda_{1}}{2} \int d^{3}x \, \sum_{k} \sum_{p} \sum_{q} \frac{e^{i\left(\boldsymbol k + \boldsymbol p + \boldsymbol q\right) \boldsymbol x}}{2)^{\frac{3}{2}}\sqrt{\omega_{k}\epsilon_{p}\epsilon_{q}}}a_{k}b_{p}b_{q},
\\
\label{eq:y:2}
& \Omega_{0}^{4,3,1} = \frac{\lambda_{1}}{2} \int d^{3}y \, \sum_{k^{'}} \frac{i\phi^{2}(y)}{\sqrt{2\omega_{k^{'}}}\omega_{k^{'}}} a_{k^{'}}^{\dagger}e^{-i \boldsymbol k^{'} \boldsymbol y},
\\
\label{eq:y:3}
& \Omega_{0}^{4,3,2} = \frac{\lambda_{0}}{4} \int d^{3}z \, \sum_{p^{'}} \sum_{q^{'}} \phi^{2}(z) \frac{i e^{-i\left(\boldsymbol p^{'} +\boldsymbol q^{'}\right)\boldsymbol z}}{2\sqrt{\epsilon_{p^{'}}\epsilon_{q^{'}}}\left(\epsilon_{p^{'}}+\epsilon_{q^{'}}\right)}b_{p^{'}}^{\dagger}b_{q^{'}}^{\dagger}.
\end{align}
\end{subequations}
The first commutator gives
\begin{equation}\label{eq:x}
    \begin{aligned} \left[H_{A}^{4,3},\Omega_{0}^{4,3,1}\right] &= \frac{i\lambda_{1}^{2}}{4} \int d^{3}x \int d^{3}y \, \sum_{k}\sum_{k^{'}}\sum_{p}\sum_{q} \frac{\phi^{2}(y)}{2^{2}}\frac{e^{i\left(\boldsymbol k + \boldsymbol p + \boldsymbol q\right)\boldsymbol x -i \boldsymbol k^{'}\boldsymbol y }}{\sqrt{\omega_{k}\omega_{k^{'}}\omega_{q}\omega_{p}}\omega_{k^{'}}} \left[a_{k}b_{p}b_{q},a_{k^{'}}^{\dagger}\right],\\
    &= \frac{i\lambda_{1}^{2}}{16} \int d^{3}x \int d^{3}y \,\sum_{p}\sum_{q} \frac{e^{i\left(\boldsymbol p +\boldsymbol q\right)\boldsymbol x}}{\sqrt{\epsilon_{p}}\epsilon_{q}M^{2}} \int \frac{d^{3}k}{\left(2\pi\right)^{3}} \, e^{i \boldsymbol k \left(\boldsymbol x -\boldsymbol y\right)}\phi^{2}(y)b_{p}b_{q},\\
    &= \frac{i\lambda_{1}^{2}}{16} \int d^{3}x \, \phi^{2}(x) \sum_{p} \sum_{q} \frac{e^{i\left(\boldsymbol p + \boldsymbol q\right)\boldsymbol x}}{M^{2}\sqrt{\epsilon_{p}\epsilon_{q}}}b_{p}b_{q}.
    \end{aligned}
\end{equation}
The second one is
\begin{equation}\label{eq:x}
    \begin{aligned} \left[\left[H_{A}^{4,3},\Omega_{0}^{4,3,1}\right],\Omega_{0}^{4,3,2}\right] &= -\frac{\lambda_{1}^{2}\lambda_{0}}{64} \int d^{3}x \int d^{3}z \, \phi^{2}(x) \phi^{2}(z), \\
    &\sum_{p}\sum_{q}\sum_{p^{'}}\sum_{q{'}} \frac{e^{i\left(\boldsymbol p + \boldsymbol q\right)\boldsymbol x-i\left(\boldsymbol p^{'}+\boldsymbol q^{'}\right)\boldsymbol z}}{2M^{2}\sqrt{\epsilon_{p}\epsilon_{q}\epsilon_{p^{'}}\epsilon_{q^{'}}}\left(\epsilon_{p^{'}}+\epsilon_{q^{'}}\right)}\left[b_{p}b_{q},b_{p^{'}}^{\dagger}b_{q^{'}}^{\dagger}\right],\\
    &=-\frac{\lambda_{1}^{2}\lambda_{0}}{64} \int d^{3}x \, \phi^{4}(x) \int \frac{d^{3}p}{\left(2\pi\right)^{3}} \frac{1}{2M^{2}\epsilon_{p}^{3}}.
    \end{aligned}
\end{equation}
As a result, 
\begin{equation}\label{eq:x}
    \begin{aligned} -\frac{1}{3}\left[\left[H_{A}^{4,3},\Omega_{0}^{4,3,1}\right],\Omega_{0}^{4,3,2}\right] = \frac{\lambda_{1}^{2}\lambda_{0}}{192} \int d^{3}x \, \phi^{4}(x) \int \frac{d^{3}p}{\left(2\pi\right)^{3}} \frac{1}{2M^{2}\epsilon_{p}^{3}}.
    \end{aligned}
\end{equation}
Including the Hermitian conjugate we have the final result for both Table \ref{tab:ch31}(b) and Table \ref{tab:ch31}(c) as,
\begin{equation}\label{eq:x}
    \begin{aligned} \frac{\lambda_{1}^{2}\lambda_{0}}{96} \int d^{3}x \, \phi^{4}(x) \int \frac{d^{3}p}{\left(2\pi\right)^{3}} \frac{1}{2M^{2}\epsilon_{p}^{3}}.
    \end{aligned}
\end{equation}
\indent For the combination in Table \ref{tab:ch31}(d), we have,
\begin{subequations}\label{eq:y}
\begin{align}
\label{eq:y:1}
&H_{A}^{4,3} = \int d^{3}x \, \frac{\lambda_{0}}{4} \phi^{2}(x) \sum_{p} \sum_{q} \frac{e^{i\left(\boldsymbol p + \boldsymbol q\right)\boldsymbol x}}{2\sqrt{\epsilon_{p}\epsilon_{q}}} b_{p}b_{q},
\\
\label{eq:y:2}
&\Omega_{0}^{4,3,1} = \int d^{3}y \,\frac{\lambda_{1}}{2} \sum_{k}\sum_{p^{'}}\sum_{q^{'}}\frac{ie^{-i\left(\boldsymbol k +\boldsymbol p^{'} +\boldsymbol q^{'}\right)\boldsymbol y}}{2^{\frac{3}{2}}\sqrt{\omega_{k}\epsilon_{p^{'}}\epsilon_{q^{'}}}\left(\omega_{k}+\epsilon_{p^{'}}+\epsilon_{q^{'}}\right)}a^{\dagger}_{k}b^{\dagger}_{p^{'}}b^{\dagger}_{q^{'}},
\\
\label{eq:y:3}
&\Omega_{0}^{4,3,2} = \int d^{3}z \, \frac{\lambda_{1}}{2}\phi^{2}(z) \sum_{k^{'}}\frac{-ie^{\boldsymbol k^{'}\boldsymbol z}}{\sqrt{2\omega_{k^{'}}}\omega_{k^{'}}}a_{k^{'}}.
\end{align}
\end{subequations}
Following similar steps as before, we first calculate the commutator:
\begin{equation}\label{eq:x}
    \begin{aligned} \left[H_{A}^{4,3},\Omega_{0}^{4,3,1}\right] &= \frac{i\lambda_{0}\lambda_{1}}{8} \int d^{3}x \int d^{3}y \, \phi^{2}(x)
    \sum_{p}\sum_{q}\sum_{k}\sum_{p^{'}}\sum_{q^{'}} \frac{e^{i\left(\boldsymbol p + \boldsymbol q\right)\boldsymbol x- i \left(\boldsymbol k + \boldsymbol p^{'} + \boldsymbol q^{'}\right)\boldsymbol y}}{2^{\frac{5}{2}}\left(\omega_{K}+\epsilon_{p^{'}}+\epsilon_{q^{'}}\right)}\\&\left[b_{p}b_{q},a_{k}^{\dagger}b_{p^{'}}^{\dagger}b_{q^{'}}^{\dagger}\right],\\
    &=\frac{i\lambda_{1}\lambda_{0}}{8} \int d^{3}x \int d^{3}y \, \phi^{2}(x) \sum_{k}\sum_{p}\sum_{q}\frac{2e^{i\left(\boldsymbol p + \boldsymbol q\right)\left(\boldsymbol x - \boldsymbol y\right)-i\boldsymbol k \boldsymbol y}}{2^{\frac{5}{2}}\sqrt{\omega_{k}}\epsilon_{p}\epsilon_{q}\left(\omega_{k}+\epsilon_{p}+\epsilon_{q}\right)}a^{\dagger}_{k},\\
    &= \frac{i\lambda_{1}\lambda_{0}}{8} \int d^{3}x \int d^{3}y \, \phi^{2}(x) \sum_{k}\sum_{p}\frac{e^{-i\boldsymbol k \boldsymbol y}}{2^{\frac{3}{2}}\epsilon_{p}^{2}\sqrt{\omega_{k}}\left(\omega_{k}+2\epsilon_{p}\right)}a^{\dagger}_{k}\\& \int \frac{d^{3}\left(p+q\right)}{\left(2\pi\right)^{3}}e^{i\left(\boldsymbol p +\boldsymbol q\right)\left(\boldsymbol x - \boldsymbol y\right)},\\
    &=\frac{i\lambda_{1}\lambda_{0}}{8} \int d^{3}x  \, \phi^{2}(x) \sum_{k}\sum_{p}\frac{e^{-i\boldsymbol k \boldsymbol x}}{2^{\frac{3}{2}}\epsilon_{p}^{2}\sqrt{\omega_{k}}\left(\omega_{k}+2\epsilon_{p}\right)}a^{\dagger}_{k}.
    \end{aligned}
\end{equation}
The total commutator then is 
\begin{equation}\label{eq:x}
    \begin{aligned} \left[\left[H_{A}^{4,3},\Omega_{0}^{4,3,1}\right],\Omega_{0}^{4,3,2}\right] &= \frac{\lambda_{1}^{2}\lambda_{0}}{16} \int d^{3}x \int d^{3}z\, \phi^{2}(z) \phi^{2}(x) \sum_{k}\sum_{p}\sum_{k^{'}}\frac{e^{i\boldsymbol k^{'}\boldsymbol z-i\boldsymbol k \boldsymbol x}}{2^{2}\epsilon_{p}^{2}\sqrt{\omega_{k}\omega_{k^{'}}}\omega_{k^{'}}\left(\omega_{k}+2\epsilon_{p}\right)}\\&\left[a_{k}^{\dagger},a_{k^{'}}\right],\\
    &=-\frac{\lambda_{1}^{2}\lambda_{0}}{64}\int d^{3}x \, \phi^{4}(x) \int \frac{d^{3}p}{\left(2\pi\right)^{3}}\,\frac{1}{M^{2}\epsilon_{p}^{2}\left(M+2\epsilon_{p}\right)}.
    \end{aligned}
\end{equation}
Hence, 
\begin{equation}\label{eq:x}
    \begin{aligned} -\frac{1}{3}\left[\left[H_{A}^{4,3},\Omega_{0}^{4,3,1}\right],\Omega_{0}^{4,3,2}\right]=\frac{\lambda_{1}^{2}\lambda_{0}}{192}\int d^{3}x \, \phi^{4}(x) \int \frac{d^{3}p}{\left(2\pi\right)^{3}}\,\frac{1}{M^{2}\epsilon_{p}^{2}\left(M+2\epsilon_{p}\right)}.
    \end{aligned}
\end{equation}
Including the Hermitian conjugate, the final contribution from the combination in Table \ref{tab:ch31}(d) is 
\begin{equation}\label{eq:x}
    \begin{aligned} \frac{\lambda_{1}^{2}\lambda_{0}}{96}\int d^{3}x \, \phi^{4}(x) \int \frac{d^{3}p}{\left(2\pi\right)^{3}}\,\frac{1}{M^{2}\epsilon_{p}^{2}\left(M+2\epsilon_{p}\right)}.
    \end{aligned}
\end{equation}
\indent The calculation of the contributions from $-\frac{1}{2}\left[\left[H_{B}^{4,3},\Omega_{0}^{4,3,1}\right],\Omega_{0}^{4,3,2}\right]$ is very similar to that of Table \ref{tab:ch31}(b) and Table \ref{tab:ch31}(c).
\subsection{Explicit Calculation of Figure \ref{fig:ch33d}}\label{sec:app33}
Consider Table 3 and in particular, the combination labelled (a). We will need the following:
\begin{subequations}\label{eq:y}
\begin{align}
\label{eq:y:1}
&H_{B}^{4,4} = \int d^{3}x \, \frac{\lambda_{0}}{2} \phi^{2}(x) \sum_{p}\sum_{q} \frac{e^{i\left(\boldsymbol p -\boldsymbol q\right)\boldsymbol x}}{2\sqrt{\epsilon_{p}\epsilon_{q}}}b_{p}^{\dagger}b_{q},
\\
\label{eq:y:2}
&\Omega_{0}^{4,4,1} = \int d^{3}y \, \lambda_{1} \phi(y) \sum_{k}\sum_{p^{'}} \frac{ie^{-i\left(\boldsymbol k + \boldsymbol p^{'}\right)\boldsymbol y}}{2\sqrt{\omega_{k}\epsilon_{p^{'}}}\left(\omega_{k}+\epsilon_{p^{'}}\right)}a_{k}^{\dagger}b_{p^{'}}^{\dagger},
\\
\label{eq:y:3}
&\Omega_{0}^{4,4,2} = \int d^{3}z \, \lambda_{1} \phi(z) \sum_{k^{'}}\sum_{q^{'}} \frac{-ie^{i\left(\boldsymbol k^{'}+\boldsymbol q^{'}\right)\boldsymbol z}}{2\sqrt{\omega_{k^{'}}\epsilon_{q^{'}}}\left(\omega_{k^{'}}+\epsilon_{q^{'}}\right)}a_{k^{'}}b_{q^{'}}.
\end{align}
\end{subequations}
The first commutator is, 
\begin{equation}\label{eq:x}
    \begin{aligned} \left[H_{B}^{4,4},\Omega_{0}^{4,4,1}\right] & =\frac{i\lambda_{0}\lambda_{1}}{2} \int d^{3}x \int d^{3}y \, \phi^{2}(x) \phi(y) \sum_{p}\sum_{q}\sum_{k}\sum_{p^{'}}\frac{e^{-i\left(\boldsymbol p - \boldsymbol q\right)\boldsymbol x - i\left(\boldsymbol k + \boldsymbol p^{'}\right)\boldsymbol y}}{4\sqrt{\epsilon_{p}\epsilon_{q}\omega_{k}\epsilon_{p^{'}}}\left(\omega_{k}+\epsilon_{p^{'}}\right)}\\&\left[b_{p}^{\dagger}b_{q},a_{k}^{\dagger}b_{p^{'}}^{\dagger}\right],\\
    & = \frac{i\lambda_{0}\lambda_{1}}{8} \int d^{3}x \int d^{3}y \, \phi^{2}(x) \phi(y) \sum_{p}\sum_{q}\sum_{k} \frac{e^{i\boldsymbol q \left(\boldsymbol x - \boldsymbol y\right)-i \boldsymbol p \boldsymbol x - i \boldsymbol k \boldsymbol y}}{\sqrt{\epsilon_{p}\omega_{k}}\epsilon_{q}\left(\omega_{k}+\epsilon_{q}\right)}b_{p}^{\dagger}a_{k}^{\dagger},\\
    & = \frac{i\lambda_{0}\lambda_{1}}{8} \int d^{3}x \int d^{3}y \, \phi^{2}(x) \phi(y) \sum_{p}\sum_{k} \frac{e^{-i \boldsymbol p \boldsymbol x - i \boldsymbol k \boldsymbol y}}{\sqrt{\epsilon_{p}\omega_{k}}\epsilon_{p}\left(\omega_{k}+\epsilon_{p}\right)}b_{p}^{\dagger}a_{k}^{\dagger} \int \frac{d^{3}q}{\left(2\pi\right)^{3}} \, e^{i\boldsymbol q \left(\boldsymbol x - \boldsymbol y\right)},\\
    & = \frac{i\lambda_{0}\lambda_{1}}{8} \int d^{3}x  \, \phi^{3}(x)  \sum_{p}\sum_{k} \frac{e^{-i \boldsymbol p \boldsymbol x - i \boldsymbol k \boldsymbol y}}{\sqrt{\epsilon_{p}\omega_{k}}\epsilon_{p}\left(\omega_{k}+\epsilon_{p}\right)}b_{p}^{\dagger}a_{k}^{\dagger},
    \end{aligned}
\end{equation}
where we have used the condition that external momenta are zero, and therefore $\left|k\right|=\left|p\right|=\left|q\right|$, which gives $\epsilon_{p}=\epsilon_{q}$.
The second commutator is 
\begin{equation}\label{eq:x}
    \begin{aligned} \left[\left[H_{B}^{4,4},\Omega_{0}^{4,4,1}\right],\Omega_{0}^{4,4,2}\right] &= \frac{\lambda_{0}\lambda_{1}^{2}}{16}\int d^{3}x \int d^{3}z \, \phi^{3}(x) \phi(z) \sum_{p}\sum_{k}\sum_{k^{'}}\sum_{q^{'}} \\
    &\frac{e^{i\left(\boldsymbol k^{'}+\boldsymbol q^{'}\right)\boldsymbol z-i\left(\boldsymbol k +\boldsymbol p\right)\boldsymbol x}}{\sqrt{\omega_{k}\omega_{k^{'}}\epsilon_{p}\epsilon_{q^{'}}}\left(\omega_{k}+\epsilon_{p}\right)\left(\omega_{k^{'}}+\epsilon_{q^{'}}\right)\epsilon_{p}}\left[b_{p}^{\dagger}a_{k}^{\dagger},a_{k^{'}}b_{q^{'}}\right],\\
    & = \frac{-\lambda_{0}\lambda_{1}^{2}}{16} \int d^{3}x \int d^{3}z \, \phi^{3}(x)\phi(z) \sum_{k}\sum_{q} \frac{e^{i\left(\boldsymbol k + \boldsymbol p\right)\left(\boldsymbol z - \boldsymbol x\right)}}{\omega_{k}\epsilon_{p}^{2}\left(\omega_{k}+\epsilon_{p}\right)^{2}},\\
    & = \frac{-\lambda_{0}\lambda_{1}^{2}}{16} \int d^{3}x  \, \phi^{4}(x) \int \frac{d^{3}k}{\left(2\pi\right)^{3}} \frac{1}{\omega_{k}\epsilon_{k}^{2}\left(\omega_{k}+\epsilon_{k}\right)^{2}}.
    \end{aligned} 
\end{equation}
Including the Hermitian conjugate, the final result for Table \ref{tab:ch33}(a) is:
\begin{equation}\label{eq:x}
    \begin{aligned} \frac{\lambda_{0}\lambda_{1}^{2}}{16} \int d^{3}x  \, \phi^{4}(x) \int \frac{d^{3}k}{\left(2\pi\right)^{3}} \frac{1}{\omega_{k}\epsilon_{k}^{2}\left(\omega_{k}+\epsilon_{k}\right)^{2}}.
    \end{aligned}
\end{equation}
\indent Next we turn to the combinations (b) and (c) of Table 3. We will need the expansions:
\begin{subequations}\label{eq:y}
\begin{align}
\label{eq:y:1}
& H_{B}^{4,4}=\lambda_{1}\int d^{3}x \, \phi(x) \sum_{k} \sum_{p} \frac{e^{i\left(\boldsymbol k -\boldsymbol p\right)\boldsymbol x}}{2\sqrt{\omega_{k}\epsilon_{p}}}b_{p}^{\dagger}a_{k},
\\
\label{eq:y:2}
&\Omega_{0}^{4,4,1}=\lambda_{1} \int d^{3}y \,\phi(y) \sum_{k^{'}}\sum_{q} \frac{ie^{-i\left(\boldsymbol k^{'} + \boldsymbol q\right)\boldsymbol y}}{2\sqrt{\omega_{k^{'}}\epsilon_{q}}\left(\omega_{k^{'}}+\epsilon_{q}\right)}a_{k^{'}}^{\dagger}b_{q}^{\dagger},
\\
\label{eq:y:3}
&\Omega_{0}^{4,4,2}=\frac{\lambda_{0}}{4} \int d^{3}z \, \phi^{2}(z) \sum_{p^{'}}\sum_{q^{'}} \frac{-ie^{i\left(\boldsymbol p^{'} + \boldsymbol q^{'}\right)\boldsymbol z}}{2\sqrt{\epsilon_{p^{'}}\epsilon_{q^{'}}}\left(\epsilon_{p^{'}}+\epsilon_{q^{'}}\right)}b_{p^{'}}b_{q^{'}}.
\end{align}
\end{subequations}
First commutator is 
\begin{equation}\label{eq:x}
    \begin{aligned} \left[H_{B}^{4,4},\Omega_{0}^{4,4,1}\right] &= i\lambda_{1}^{2}\int d^{3}x \int d^{3}y \, \phi(x)\phi(y) \sum_{k}\sum_{p}\sum_{k^{'}}\sum_{q} \frac{e^{i\left(\boldsymbol k - \boldsymbol p\right)\boldsymbol x - i\left(\boldsymbol k^{'}+\boldsymbol q\right)\boldsymbol y}}{4\sqrt{\omega_{k}\omega_{k^{'}}\epsilon_{p}\epsilon_{q}}\left(\omega_{k^{'}}+\epsilon_{q}\right)}\\
    &\left[a_{k}b_{p}^{\dagger},a{k^{'}}^{\dagger}b_{q}^{\dagger}\right],\\
    &=\frac{i\lambda_{1}^{2}}{4}\int d^{3}x \int d^{3}y \, \phi(x)\phi(y) \sum_{k}\sum_{p}\sum_{q} \frac{e^{i\left(\boldsymbol x - \boldsymbol y\right)\boldsymbol k - i\boldsymbol p \boldsymbol x -i\boldsymbol q \boldsymbol y}}{\omega_{k}\sqrt{\epsilon_{p}\epsilon_{q}}\left(\omega_{k}+\epsilon_{q}\right)}b_{p}^{\dagger}b_{q}^{\dagger},\\
    &=\frac{i\lambda_{1}^{2}}{4} \int d^{3}x \, \phi^{2}(x) \sum_{p} \sum_{q} \frac{e^{-i\left(\boldsymbol p + \boldsymbol q\right) \boldsymbol x }}{\omega_{p}\sqrt{\epsilon_{p}\epsilon_{q}}\left(\omega_{p}+\epsilon_{q}\right)}b_{p}^{\dagger}b_{q}^{\dagger}.
    \end{aligned}
\end{equation}
Again, we use the condition that external momenta are zero, and rewrite $\omega_{k}$ as $\omega_{p}$, since $\left|k\right|=\left|p\right|$.
The second commutator gives:
\begin{equation}\label{eq:x}
    \begin{aligned} \left[\left[H_{B}^{4,4},\Omega_{0}^{4,4,1}\right],\Omega_{0}^{4,4,2}\right]&=\frac{\lambda_{1}^{2}\lambda_{0}}{16} \int d^{3}x \int d^{3}z \, \phi^{2}(x)\phi^{2}(z) \sum_{p}\sum_{q}\sum_{p^{'}}\sum_{q^{'}}\\&\frac{e^{-i\left(\boldsymbol p + \boldsymbol q\right)\boldsymbol x + i\left(\boldsymbol p^{'}+\boldsymbol q^{'}\right)\boldsymbol z}}{2\omega_{p}\sqrt{\epsilon_{p}\epsilon_{q}\epsilon_{p^{'}}\epsilon_{q^{'}}}\left(\epsilon_{p^{'}}+\epsilon_{q^{'}}\right)\left(\omega_{p}+\epsilon_{q}\right)}\left[b_{p}^{\dagger}b_{q}^{\dagger},b_{p^{'}}b_{q^{'}}\right],\\
    & = \frac{-\lambda_{1}^{2}\lambda_{0}}{16}\int d^{3}x \int d^{3}z \, \phi^{2}(x) \phi^{2}(z) \sum_{p}\sum_{q}\frac{e^{i\left(\boldsymbol p + \boldsymbol q\right)\left(\boldsymbol z - \boldsymbol x\right)}}{\omega_{p}\epsilon_{p}\epsilon_{q}\left(\epsilon_{p}+\epsilon_{q}\right)\left(\omega_{p}+\epsilon_{q}\right)},\\
    & = \frac{-\lambda_{1}^{2}\lambda_{0}}{32} \int d^{3}x \, \phi^{4}(x) \int \frac{d^{3}p}{\left(2\pi\right)^{3}} \, \frac{1}{\omega_{p}\epsilon_{p}^{3}\left(\omega_{p}+\epsilon_{p}\right)}.
    \end{aligned}
\end{equation}
Including the Hermitian conjugate, the final result for both Table \ref{tab:ch33}(b) and Table \ref{tab:ch33}(c) is:
\begin{equation}\label{eq:x}
    \begin{aligned} \frac{\lambda_{1}^{2}\lambda_{0}}{32} \int d^{3}x \, \phi^{4}(x) \int \frac{d^{3}p}{\left(2\pi\right)^{3}} \, \frac{1}{\omega_{p}\epsilon_{p}^{3}\left(\omega_{p}+\epsilon_{p}\right)}.
    \end{aligned}
\end{equation}



\begin{thebibliography}{99}

\bibitem{Wilson:1971bg}
K.~G.~Wilson,
``Renormalization group and critical phenomena. 1. Renormalization group and the Kadanoff scaling picture,''
Phys. Rev. B \textbf{4} (1971), 3174-3183
doi:10.1103/PhysRevB.4.3174 

\bibitem{Wilson:1971dh}
K.~G.~Wilson,
``Renormalization group and critical phenomena. 2. Phase space cell analysis of critical behavior,''
Phys. Rev. B \textbf{4} (1971), 3184-3205
doi:10.1103/PhysRevB.4.3184

\bibitem{Polchinski:1983gv}
J.~Polchinski,
``Renormalization and Effective Lagrangians,''
Nucl. Phys. B \textbf{231} (1984), 269-295
doi:10.1016/0550-3213(84)90287-6

\bibitem{Appelquist:1974tg}
T.~Appelquist and J.~Carazzone,
``Infrared Singularities and Massive Fields,''
Phys. Rev. D \textbf{11} (1975), 2856
doi:10.1103/PhysRevD.11.2856

\bibitem{Georgi:1994qn}
H.~Georgi,
``Effective field theory,''
Ann. Rev. Nucl. Part. Sci. \textbf{43} (1993), 209-252
doi:10.1146/annurev.ns.43.120193.001233

\bibitem{Manohar:1996cq}
A.~V.~Manohar,
``Effective field theories,''
Lect. Notes Phys. \textbf{479} (1997), 311-362
doi:10.1007/BFb0104294
[arXiv:hep-ph/9606222 [hep-ph]]

\bibitem{Rothstein:2003mp}
I.~Z.~Rothstein,
``TASI lectures on effective field theories,''
[arXiv:hep-ph/0308266 [hep-ph]].

\bibitem{Symanzik:1981wd}
K.~Symanzik,
``Schrodinger Representation and Casimir Effect in Renormalizable Quantum Field Theory,''
Nucl. Phys. B \textbf{190} (1981), 1-44
doi:10.1016/0550-3213(81)90482-X

\bibitem{Pi:1987df}
S.~Y.~Pi and M.~Samiullah,
``Renormalizability of the Time Dependent Variational Equations in Quantum Field Theory,''
Phys. Rev. D \textbf{36} (1987), 3128
doi:10.1103/PhysRevD.36.3128

\bibitem{Minic:1994ff}
D.~Minic and V.~P.~Nair,
``Wave functionals, Hamiltonians and the renormalization group,''
Int. J. Mod. Phys. A \textbf{11} (1996), 2749-2764
doi:10.1142/S0217751X96001334
[arXiv:hep-th/9406074 [hep-th]]

\bibitem{Glazek:1994qc}
S.~D.~Glazek and K.~G.~Wilson,
``Perturbative renormalization group for Hamiltonians,''
Phys. Rev. D \textbf{49} (1994), 4214-4218
doi:10.1103/PhysRevD.49.4214

\bibitem{Gubankova:1997mq}
E.~L.~Gubankova and F.~Wegner,
``Flow equations for QED in the light front dynamics,''
Phys. Rev. D \textbf{58} (1998), 025012
doi:10.1103/PhysRevD.58.025012
[arXiv:hep-th/9710233 [hep-th]].

\bibitem{Alexanian:1998wu}
G.~Alexanian and E.~F.~Moreno,
``On the renormalization of Hamiltonians,''
Phys. Lett. B \textbf{450} (1999), 149-157
doi:10.1016/S0370-2693(99)00136-7
[arXiv:hep-th/9811158 [hep-th]].

\bibitem{Balasubramanian:2011wt}
V.~Balasubramanian, M.~B.~McDermott and M.~Van Raamsdonk,
``Momentum-space entanglement and renormalization in quantum field theory,''
Phys. Rev. D \textbf{86} (2012), 045014
doi:10.1103/PhysRevD.86.045014
[arXiv:1108.3568 [hep-th]].

\bibitem{Headrick:2018ctr}
M.~Headrick,
``Entanglement in Field Theory and Holography,''
PoS \textbf{TASI2017} (2018), 012
doi:10.22323/1.305.0012

\bibitem{Ryu:2006ef}
S.~Ryu and T.~Takayanagi,
``Aspects of Holographic Entanglement Entropy,''
JHEP \textbf{08} (2006), 045
doi:10.1088/1126-6708/2006/08/045
[arXiv:hep-th/0605073 [hep-th]].






\end{thebibliography}
\end{document}